\RequirePackage{snapshot}
\documentclass[useAMS,usenatbib, fleqn]{mn2e}

\usepackage{graphicx}

\usepackage{ifpdf}

\usepackage{times}

\usepackage{fixltx2e}

\usepackage{hyperref}
\hypersetup{colorlinks=true,citecolor=blue,linkcolor=black,filecolor=black,runcolor=black}

\usepackage{booktabs}

\usepackage{aas_macros}

\ifpdf
	\usepackage{epstopdf}
\fi

\usepackage{amsmath}
\usepackage{amssymb}

\title[Minimum planet mass detectable in discs]{The minimum mass of  detectable planets in protoplanetary
discs and the derivation of planetary masses from high resolution observations}
\author[G. P. Rosotti et al.]{Giovanni P. Rosotti\thanks{E-mail:
rosotti@ast.cam.ac.uk}$^{1}$, Attila Juhasz$^{1}$, Richard A. Booth$^{1}$, Cathie J. Clarke$^{1}$\\
$^{1}$Institute of Astronomy, University of Cambridge, Madingley Road, Cambridge CB3 0HA, UK}

\begin{document}

\date{Accepted . Received ; in original form }

\pagerange{\pageref{firstpage}--\pageref{lastpage}} \pubyear{2016}

\maketitle

\label{firstpage}

\begin{abstract}

We investigate the minimum planet mass that produces observable signatures in infrared scattered light and submm continuum images and demonstrate how these images can be used to {\it measure} planet masses to within a factor of about two. To this end we perform multi-fluid gas and dust simulations of discs containing low mass planets, generating simulated observations at $1.65 \mu$m, $10 \mu$m and $850 \mu$m. We show that the minimum planet mass that produces a detectable signature is $\sim 15 M_\oplus$: this value is strongly dependent on disc temperature and changes slightly with wavelength (favouring the submm). We also {confirm previous results} that there is a minimum planet mass of $\sim 20 M_\oplus$ that produces a pressure maximum in the disc: only planets above this threshold mass generate a dust trap that can eventually create a hole in the submm dust. Below this mass, planets produce annular enhancements in dust outward of the planet and a reduction in the vicinity of the planet. These features are in steady state and can be understood in terms of variations in the dust radial velocity, imposed by the perturbed gas pressure radial profile, analogous to a traffic jam. We also show how planet masses can be derived from structure in scattered light and sub-mm images. We emphasise that simulations with dust need to be run over thousands of planetary orbits so as to allow the gas profile to achieve a steady state and caution against the estimation of planet masses using gas only simulations.
\end{abstract}

\begin{keywords}
hydrodynamics, protoplanetary discs, planet-disc interactions, submillimetre: planetary systems
\end{keywords}

\section{Introduction}

Planets form in the dense, rotationally flattened structures of dust and gas called ``proto-planetary disc'' (see \citealt{WilliamsCieza} for a review). Despite the rapid  expansion of our knowledge of exoplanets around main sequence stars \citep{1995exoplanet}, little is known about the putative planets that may already be present in such discs at an age of $ < 10$ Myr. The Atacama Large Millimeter/Submillimeter Array (ALMA), which is approaching its full capability, can reach the unprecedented resolution and sensitivity necessary for the detailed characterisation of such discs and may also
be able to detect direct signatures of embedded protoplanets (see for
example the  pattern of bright and dark rings observed in HL Tau during a science verification campaign which has been interpreted as being of
planetary origin; \citealt{hltau}). The latest near-infrared (NIR) instruments on 8\,m-class telescopes (e.g. Spectro-Polarimetric High-contrast Exoplanet REsearch (SPHERE) on the Very Large Telescope (VLT) or Gemini Planet Imager (GPI) on the Gemini Telescope) provide an opportunity
to search for planetary signatures in discs via scattered light imaging at similar resolution as ALMA in the sub-millimetre. In the near future similar resolution will be available in the thermal mid-infrared (MIR) when the 30--40m class telescopes (e.g. the European Extremely Large Telescope, E-ELT) will come online.

%

 It is very tempting to connect the presence of rings and gaps in 
protoplanetary disc images (which might be commonplace, see \citealt{2016arXiv160105182Z}) to the presence of planets. According to conventional core accretion
models \citep{Pollack96}, however, giant planets are an unexpected outcome at the
very young age ascribed to  HL Tau  ($10^5$ yr; \citealt{KenyonHartmann95}). It is however
currently unclear whether this indeed involves a revision of planet
formation mechanisms and timescales or whether the structures seen
in the image are not planetary in origin. Our focus here however is
not to add to the debate on this particular system but to present a more
generic exploration of the detectability of planetary signatures via
submm and scattered light imaging. 

It is often assumed  that only giant planets can create structures in  discs prominent enough to be observed (e.g., \citealt{varniere20062,2013A&A...549A..97R,2014A&A...572L...2R}). Indeed, it is well known that giant planets create gaps in the \textit{gas} disc \citep{GapOpening}, partially depleting the surface density in a region as wide as a few times its Hill radius. 
Planets with a mass smaller than a critical value, which is typically somewhat lower than the mass of Jupiter, are not able to open { significant} gaps \citep{CridaOpeningCriterion}, and one might then think that these objects would not produce observable features in discs. 
In this case it would not be possible to put observational constraints on the formation conditions of the large population of super-Earths
discovered by Kepler around
main sequence stars, objects now believed to constitute  the most abundant population of planetary
objects \citep{Howard2012}. This is particularly unfortunate because whereas it is clear
that gas giant planets {\it must} form at an evolutionary stage
when disc gas is still abundantly present, it is unclear whether super-Earths
originate at similarly early times. Conventionally, terrestrial planet
formation is often ascribed to later eras (age of $\sim 100$ Myr) when the
disc gas has long since dissipated and planet building proceeds via
planetesimal collisions { and giant impacts among embryos} in a gas poor environment \citep{2014prpl.conf..595R}. On the other
hand, the conventional core accretion scenario for gas giant planet
formation \citep{Pollack96} envisages the formation of rock cores within the protoplanetary
disc which only convert to gas giants if they attain a critical mass in excess
of 10-20 Earth masses while gas is still abundantly present. An alternative model
for super-Earth formation would therefore involve the assembly of rock cores
during the disc phase, in the case that cores did not achieve
criticality during the gas rich phase. {The core growth rate, and therefore the dichotomy that we observe for example in the Solar System between terrestrial and giant planets, might be due to the location in the disc with respect to the water snow line \citep{2015Icar..258..418M}.} A conceptually distinct scenario for
super-Earth formation instead invokes photoevaporation from the central
star (on a timescale of $\sim 100$ Myr) to erode the gas envelope of a gas giant planet
formerly produced in the protoplanetary disc \citep{2013ApJ...775..105O}.

 Here we explore whether planetary signatures can potentially be detected
in the low mass (super-Earth) regime. This issue has not received
 much attention to date since the impossibility of gap opening
at masses much below a Jupiter mass ($\sim 300 M_\oplus$) implies
that low mass planets have only a modest effect on disc 
{\it gas} distributions. Nevertheless a number of studies have indicated
that planet induced signatures are stronger in disc {\it dust} than
in the gas, 
 on account of the tendency of drag-coupled dust  to collect in
pressure maxima in the disc \citep{metersizedproblem}. Such an effect
has been shown to enhance the observability of 
structures produced by more massive
planets, as dust becomes trapped at the outer edge of disc gaps 
( e.g. \citealt{Rice2006,Zhu2012,OwenRadPress}).

\citet{paardekooper2004} were the first to simulate dust gaps opened by low mass planets, finding  gaps opening at resonances. Their  simulated ALMA images
demonstrated the capability of ALMA to observe such structures: they
found that a 0.05 $M_j$ (15 $M_\oplus$) planet opens a gap in the dust 
while a 0.01 $M_j$ (3 $M_\oplus$) does not, but do not elaborate further on
the detectability threshold.
In an extensive work, \citet{zhu2014} showed  that even a planet with a mass as little as $8 \ M_\oplus$ can affect the dust surface density, creating  a double gapped structure. However the observational consequences are not explored in this paper. Moreover  this work considers only inviscid discs and therefore cannot
explore the effect of viscously driven inflow in the gas in modifying the resultant dust structures.
\citet{Dong2015}  focussed on giant planets and their impact on transitional discs, but they also simulate a 0.2 $M_J$ planet which they find is able to open a gap in the dust. They also find that such a low mass planet is not able to affect the spectral energy distribution (SED). To the best of our knowledge their work is the only one that, when exploring the observational consequences, is not restricted to ALMA wavelengths but also includes simulated near-infrared (NIR) images. However no study is made of systematically lowering the planet mass until the observed signatures disappear.
In a series of papers, \citet{2010A&A...518A..16F} explored  the impact of planets on the dust distributions, but in  creating synthetic ALMA observations \citep{2012A&A...547A..58G} only the gas giant regime is explored.
Even more recently, the HL Tau observations \citep{hltau} prompted other works which have focussed on explaining the ring structure that was observed. \citet{Dipierro2015} interpreted the image using three planets, finding a best fit with masses of 0.2, 0.27 and 0.55 $M_j$. \citet{Picogna2015} also interpreted  the HL Tau image in terms  of planets, but in this case only two planets are invoked (with best fit masses 0.07 and 0.35 $M_j$). The authors also commented that, depending on the disc parameters, they find observable gaps even assuming masses of 10 and 20 $M_\oplus$. { Finally, \citet{2016ApJ...818...76J} also interpreted the image using three planets, but with best fit masses of 0.35, 0.17 and 0.26 $M_j$.}

The above  works  have shown convincingly that there is a prospect for observing low mass planets in discs. However, as we have highlighted, no work so far has directly established what is the minimum planet mass that creates observable features at high resolution with current instruments (or those planned for the near future).  In addition, most of the effort has concentrated on ALMA and very little attention has been dedicated to scattered light images, which however have the same resolution as ALMA. Finally, to the best of our knowledge, the literature contains  no predictions of this kind for MIR thermal images; with the upcoming generation of 30m class telescopes, images at these wavelengths will in a few years have the same resolution as existing sub-mm and NIR instruments. The goal of this paper is to remedy this omission by deriving a threshold for the observability of a planetary gap in 
protoplanetary discs at NIR, MIR and submm wavelengths, as well as studying the dependence of this threshold on disc properties.



This paper is structured as follows. We present our method in section \ref{sec:method} and the results from the multi-fluid simulations in section \ref{sec:results_hydro}. We then present the simulated observations in section \ref{sec:sim_obs}. Section \ref{sec:discussion} discusses our results and we draw our conclusions in section \ref{sec:concl}.



\section{Numerical method and initial conditions}

\label{sec:method}

Our methodology consists of  running 2D multi-fluid gas dust simulations of planet disc interactions. We then post-process the simulations by calculating synthetic observations at three different wavelengths (NIR, MIR and sub-mm).

\subsection{Gas and dust dynamics}
We run 2D multi fluid simulation where we evolve at the same time the dust and the gas using the \textsc{fargo-3d} code \citep{2015Natur.520...63B}. \textsc{fargo} uses the \textsc{zeus} numerical algorithm \citep{Zeus}. The algorithm is well tested and it has been used many times for proto-planetary disc studies (see \citealt{ValBorro2006} for an algorithmic comparison). We extended the code to include dust, approximating the dust as a pressureless fluid that is coupled to the gas via linear drag forces. We have neglected feedback from the dust onto the gas. These approximations are valid for low dust-to-gas ratios and particles with Stokes number (see next paragraph) $\mathrm{St} <1 $ \citep{Garaud2004}.

The equation for the dust velocity, $\vec{v}_d$, is given by
\begin{equation}
\frac{\mathrm{d}\vec{v_d}}{\mathrm{d}t} + \vec{v_d}\cdot \nabla \vec{v_d}  = - \frac{1}{t_s}(\vec{v}_d - \vec{v}_g(t)) + \vec{a}_d, \label{Eqn:DustAccel}
\end{equation}
where $\vec{v}_g$ is the gas velocity, $\vec{a}_d$ are the non-drag accelerations felt by the dust, and $t_s$ is the stopping time. It is common to introduce the dimensionless stopping time, also called Stokes number, defined as $St=t_s \Omega_k$, where $\Omega_k$ is the keplerian angular velocity at the given location in the disc. We solve this equation along with the continuity equation $\tfrac{\partial \Sigma_d}{\partial t} + \nabla \cdot (\Sigma_d \vec{v}_d) = 0$, using the \textsc{zeus} algorithm, in which the forces (source step, RHS of Eqn.~\ref{Eqn:DustAccel}) are evaluated prior to the advection (transport step, $\vec{v_d}\cdot \nabla \vec{v_d}$ term). The transport step for the dust is identical to that of the gas. For details, see \citet{Zeus}, \citet{Masset2000} and references therein.

The source step for the dust is evaluated semi-implicitly, using the analytical solutions available for the simple form of the drag law. We take into account the change in gas velocity over the time-step by approximating $\vec{v}_g$ as ${\vec{v}_g(t') = \vec{v}_g(t) + \vec{a}_g \times (t' - t)}$ throughout the time step, where $\vec{a}_g$ is the acceleration calculated explicitly ({ that is, at time $t$}) during the source step for the gas. The dust velocity at $t + \Delta t$ is given by
\begin{align}
\vec{v}_d(t + \Delta t) &= \vec{v}_d(t) \exp(-\Delta t/t_s)  + \vec{a}_g \Delta t\nonumber\\
&\quad + \left[\vec{v}_g(t) + (\vec{a}_d-\vec{a}_g)t_s \right](1 - \exp(-\Delta t/t_s)),
\end{align}
which reproduces the explicit update when $\Delta t \ll t_s$ and the short friction time limit when $\Delta t \gg t_s$. To see this, consider $\vec{a}_d = \vec{g}$ and $\vec{a}_g = - \tfrac{\nabla P}{\Sigma_g} + \vec{g}$, where $\vec{g}$ is the gravitational acceleration and $P$ is the gas pressure. For $\Delta t \gg t_s$
\begin{equation}
\vec{v}_d(t + \Delta t) \rightarrow \vec{v}_g(t + \Delta t) + t_s \tfrac{\nabla P}{\Sigma_g},
\end{equation}
i.e. the short friction time limit.

\begin{figure}
\centering
\includegraphics[width=\columnwidth]{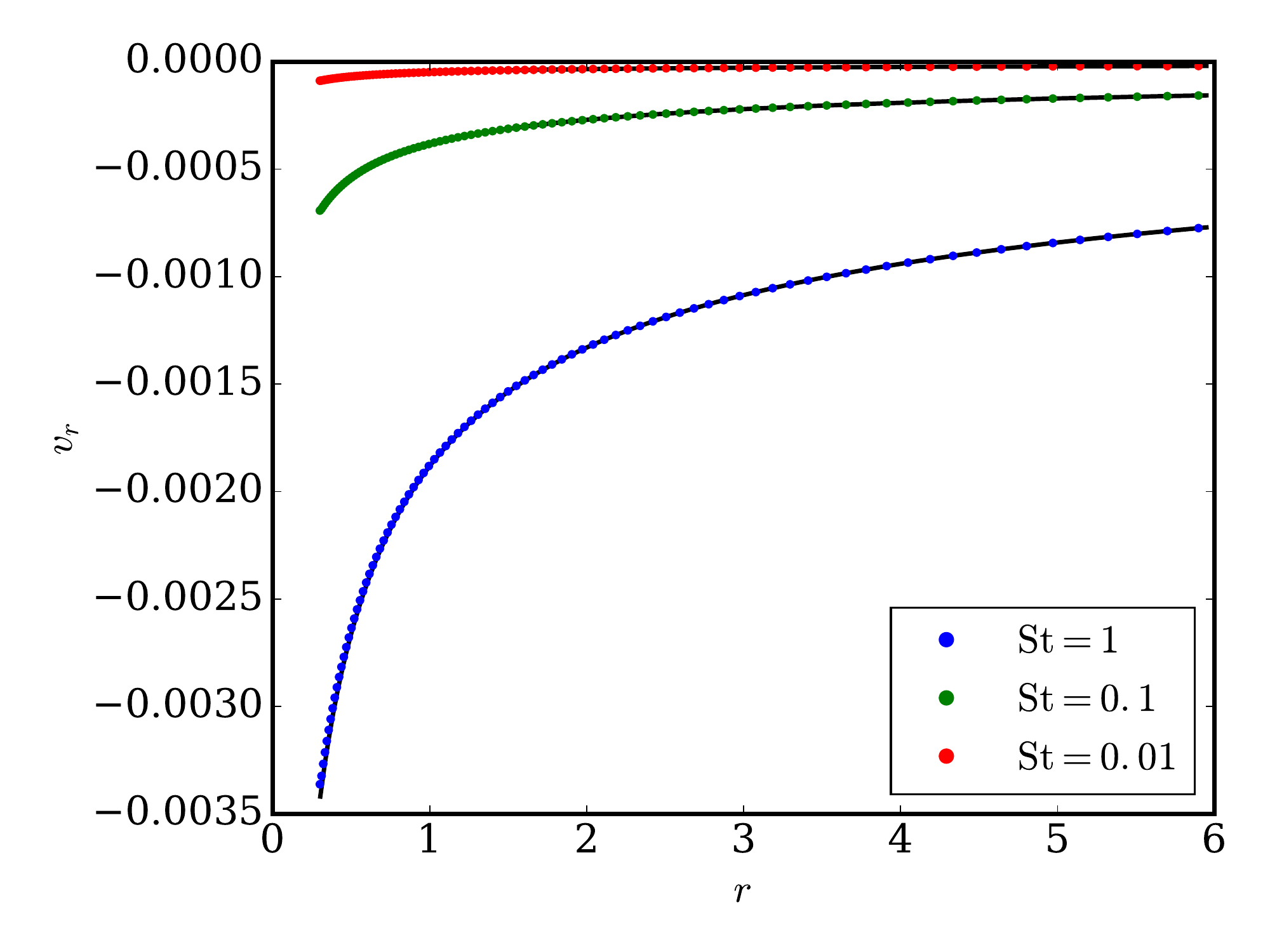}
\caption{Radial velocity of the dust in a 2D disc for a range of Stokes Numbers. The velocity from the simulation (points) agrees well with the analytical solution (solid lines, \citealt{TakeuchiLin2002}).}
\label{fig:Drift}
\end{figure}

The timestep, $\Delta t$, is limited by both the timestep of the gas and via a CFL-like criterion for the dust, ${\Delta t = C \Delta x / \max(|v_d|,|v_d - v_g|)}$, where we use a conservative value of $C = 0.44$ for both the gas and dust. We have verified the technique using a range of tests. For example, Fig.~\ref{fig:Drift} shows the code achieves good agreement between the radial velocity obtained in a low resolution  ($N_r\times N_\phi = 350 \times 580$) 2D simulation and the analytical solution \citep{TakeuchiLin2002}.

Additionally, mass diffusion was added to the surface density in the dust using a Schmidt number $\mathrm{Sc} = 1$. The diffusive mass flux, $\vec{F}_D = - (\nu/\mathrm{Sc}) \Sigma_g \nabla(\Sigma_D/\Sigma_G)$, where $\nu$ is the turbulent viscosity \citep{ClarkeDiffusion}, was included by adding an effective diffusive velocity $\vec{v}_D = \vec{F_D} / \Sigma_D$ to the velocity at which the dust is advected during the transport step. 

\subsection{Initial conditions and parameters}
\label{sec:ic}

For the viscosity we use the $\alpha$ prescription of \citet{shakura_1973}   assuming $\alpha = 10^{-3}$. 
We use 2D cylindrical coordinates and dimensionless units in which the orbital radius of the planet ($r_p$) is at unity, the unit of mass is that of the central star, while the unit time is the inverse of the Kepler-frequency of the planet. 
The inner radial boundary of our grid is at 0.5 $r_p$ and the outer boundary at $3 r_p$; we use non reflecting boundary conditions at both boundaries. The resolution is 450 and 1024 uniformly spaced cells in the radial\footnote{While this gives a different relative resolution across the grid, the limited range in radii of our grid does not make the effect significant.} and azimuthal direction, respectively. { We let the disc relax for a time $t_0$ of 10 orbits before introducing the planet, and then, to avoid numerical artefacts, we increase the mass of the planet from zero to the final one over a time $t_\mathrm{taper}$ of 20 orbits, according to the formula $\sin^2 (\pi/2 \times (t-t_0)/t_\mathrm{taper})$. The planet is kept on a circular orbit whose orbital parameters are not allowed to vary.} The surface density profile is given by:
\begin{equation}
\Sigma(R)=\Sigma_0 \frac{r_p}{R},
\label{eq:sigma}
\end{equation}
where the value of the normalisation constant $\Sigma_0$ is arbitrary as far as the dynamics is concerned. In our fiducial model the disc has an aspect ratio $H/R$ at the location of the planet of 0.05. The aspect ratio varies in a power-law fashion with a flaring index of 0.25. The value of the aspect ratio is particularly important as it controls the strength of the pressure forces, which are responsible (see section \ref{sec:gassigma}) for closing the planetary gap. For this reason we also run models with a different value of the aspect ratio. We chose values of 0.025 (``cold'') and 0.1 (``hot'') that encompass the possible range of variation in real proto-planetary discs.

For the fiducial model, we { run simulations with different planet masses, considering} values of $8,12,20,60,120 \ M_\oplus$. { Note that as far as the dynamics is concerned only the ratio of the planet mass to the star mass matters; the planet masses we quote assume a central mass of $1 \ M_\odot$, and need to be properly rescaled with the star mass.} When varying the aspect ratio of the disc, we also run models for additional values of the planet mass, as the aspect ratio controls how massive a planet must be to significantly affect the surface density of the disc. For the ``cold'' disc case ($H/R=0.025$) we also run models for a mass of $2.5 \ M_\oplus$, and for the ``hot'' disc case for masses of $96,160, 480 \ M_\oplus$.

Along with the gas dynamics we integrate in time the evolution of the dust. To keep our simulations scale-free, we fix the Stokes number $St$ of the grains rather than their physical size. Once the physical scales of the disc have been chosen, it is then possible to convert our Stokes numbers to real sizes as we explain in the next section. We use 5 dust sizes, with Stokes numbers (logarithmically spaced) ranging from $2 \times 10^{-3}$ to 0.2.


\subsection{Radiative transfer}

To investigate the detectability of dust gaps opened by low mass planets we calculate images at various wavelengths using the 3D radiative transfer code
\textsc{radmc-3d}\footnote{http://www.ita.uni-heidelberg.de/~dullemond/software/radmc-3d/}. In the radiative transfer calculation we use a 3D spherical
mesh with N$_r$=256, N$_\theta$=180, N$_\phi$=384 grid points in the radial, poloidal and azimuthal direction, respectively. For the radial direction we use a logarithmic grid extending from 1\,au to 100\,au while an arithmetic grid is used for the angular coordinates. To ensure that we resolve the disc properly in the vertical direction 
we place $N_\theta=\{10,160,10\}$ points in the $\left[0,\pi/2-\theta_0\right]$,  $\left[\pi/2-\theta_0,\pi/2+\theta_0\right]$, $\left[\pi/2+\theta_0,\pi\right]$
intervals, respectively, where $\theta_0$ = 5$H/R$, $H/R$ being the aspect ratio of the disc taken at the outer edge of the disc. 

The disc density distribution { for a given grain size $a$} is assumed to be 
\begin{equation}
\rho_a(R,z,\phi) = \frac{\Sigma_a(R, \phi)}{\sqrt{2\pi}H(R)}\exp{\left(-\frac{z^2}{2h(R)^2}\right)}
\label{eq:disc_density}
\end{equation}
where $\Sigma_a(R,\phi)$ is the  dust surface density, $R=r\sin{\theta}$ and $z=r\cos{\theta}$ and
$H(R)$ is the pressure scale height. For the latter we use the same value, as a function of radius, as is assumed in the hydrodynamic simulations.
The dust surface density of the disc is taken from the hydrodynamic simulation in the following way. First we take both the gas and dust surface density from
the hydrodynamic simulations and { bilinearly} interpolate them to the spatial grid used in the radiative transfer calculations. We extrapolate the disc both inwards and
outwards of the hydrodynamic computational domain\footnote{In particular, we do not consider the first 6 cells of hydrodynamical computational domain, which are affected by the inner boundary condition and show an artificial dust accumulation} if it is necessary assuming that  $\Sigma(R)\propto R^{-1}$. 
Then we calculate the gas density structure using Eq.\,(\ref{eq:disc_density}). { As our hydrodynamics simulations are not 3D, we did not include dust settling or iterated to reach hydrostatic equilibrium. Including these effects properly would require including also other 3D effects which are potentially more important, which goes beyond the scope of this paper (see the discussion in section \ref{sec:limitation}). Neglecting settling is particularly relevant for the sub-mm dust; however we note that, given that it is optically thin, observations mostly probe the dust surface density, and at sub-mm wavelengths are not really sensitive to the details of the vertical structure. In addition, while settling might change the global temperature of the disc, this would result only in a global change of the surface brightness, whereas here we are interested in the sub-structure present in the image. The same argument can be applied to another inconsistency that stems from our approach, namely that the temperature assumed for the hydro calculation might be different from the one computed by \textsc{radmc-3d}. Also, we remark that such inconsistencies always arise when using parametric models; the benefit is that they allow us to set up a controlled environment.}

The dimensionless surface density provided by \textsc{fargo-3d} is converted
to dimensional form taking the planetary orbital radius to be 30\,AU and normalising the density such that the gas mass in the disc is 
0.01\,M$_\odot$ { within 100 AU}. { The corresponding normalisation of the gas surface density at 1 AU is $\sim 300 \ \mathrm{g} \ \mathrm{cm}^{-2}$. For what concerns the dust, we normalise the density so that the initial dust to gas ratio is $10^{-2}$. We use 10 logarithmically spaced grain size bins between 0.1\,$\mu$m and 1\,mm and assume that the dust grain size distribution follows $dN/da \propto a^{-3.5}$}. 

Then for a given dust grain size we calculate the Stokes number in the disc midplane, assuming that the particles are in the Epstein regime:
\begin{equation}
St=t_s \Omega=\frac{a \rho_d \Omega}{\rho_g c_s}=\frac{a \rho_d}{\Sigma_g},
\end{equation}
where $a$ is the size of the particles, $\rho_d$ is the bulk density of the dust, which we assume to be 3.6\,g/cm$^3$, { $\rho_g$ is the density of the gas and $\Sigma_g$ the surface density of the gas. To arrive at the last expression we have used the fact that $\Sigma_g = \rho_g H$ and $H=c_s/\Omega$; note that in these equations we have neglected factors of order unity}. At any given location in the disc we {perform a  linear interpolation in Stokes number to compute the dust surface density, starting from the results of our hydrodynamical simulations}. When the Stokes number is smaller than the smallest one we have  in the multi-fluid simulation we assume that the dust follows the gas. We note that with our parameters the typical Stokes number of 1\,mm particles at 30 AU is $\sim 0.07$ and $\sim 0.17$ at 100 AU, which ensures that we have enough information from the multi-fluid simulations to reconstruct the surface density of those particles.

The mass absorption coefficients of the dust grains are calculated 
with Mie-theory using the optical constants of astronomical silicates \citep{Weingartner2001}. The radiation field of the central star is modelled
with blackbody emission and the star is assumed to have parameters, typical for a Herbig Ae star\footnote{The spectral type and luminosity of the
central star (e.g. HAe vs TTs) affects only the absolute surface brightness of the disc but does not influence the morphology of the images. The choice 
of using a Herbig Ae stellar model is motivated by the fact that most of the sources observed by current instruments in the NIR like SPHERE or GPI are  Herbig Ae stars due to sensitivity limitations.},  M$_\star$=2\,M$_\odot$, T$_{\rm eff}$ = 9500\,K, R$_\star$=2.5\,R$_\odot$. 

As a first step, we calculate the temperature of the dust with a thermal Monte Carlo simulation, then we calculate images at 1.65\,$\mu$m, 10\,$\mu$m and 880\,$\mu$m taking the disc inclination to be 10$^\circ$. We use 1.6$\cdot 10^8$ photons and $9\cdot 10^7$ photons for the thermal Monte Carlo simulations and for the image calculations, respectively.

\section{Results from the gas and dust dynamics simulation}

\label{sec:results_hydro}

\subsection{Gas surface density}

\label{sec:gassigma}

\begin{figure}
\includegraphics[width=\columnwidth]{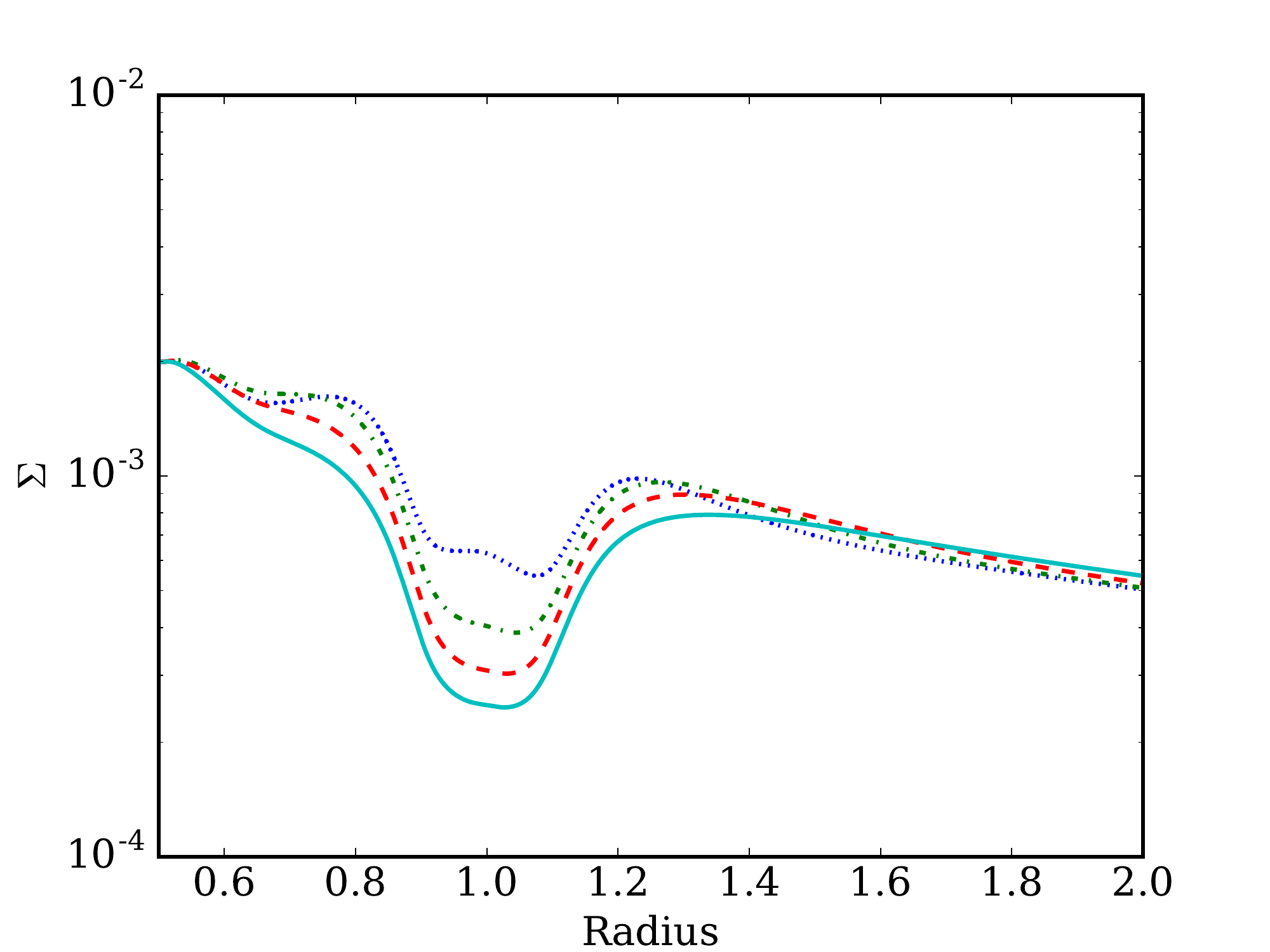}
\caption{Azimuthally averaged gas surface density for the fiducial model for a planet mass of 60 $M_\oplus$. Different lines correspond to different snapshots from the simulations: 150 orbits (blue dotted line), 400 orbits (green dotted-dashed line), 1000 orbits (red dashed line) and finally 3000 orbits (solid cyan line).}
\label{fig:gas_t}
\end{figure}

\begin{figure}
\includegraphics[width=\columnwidth]{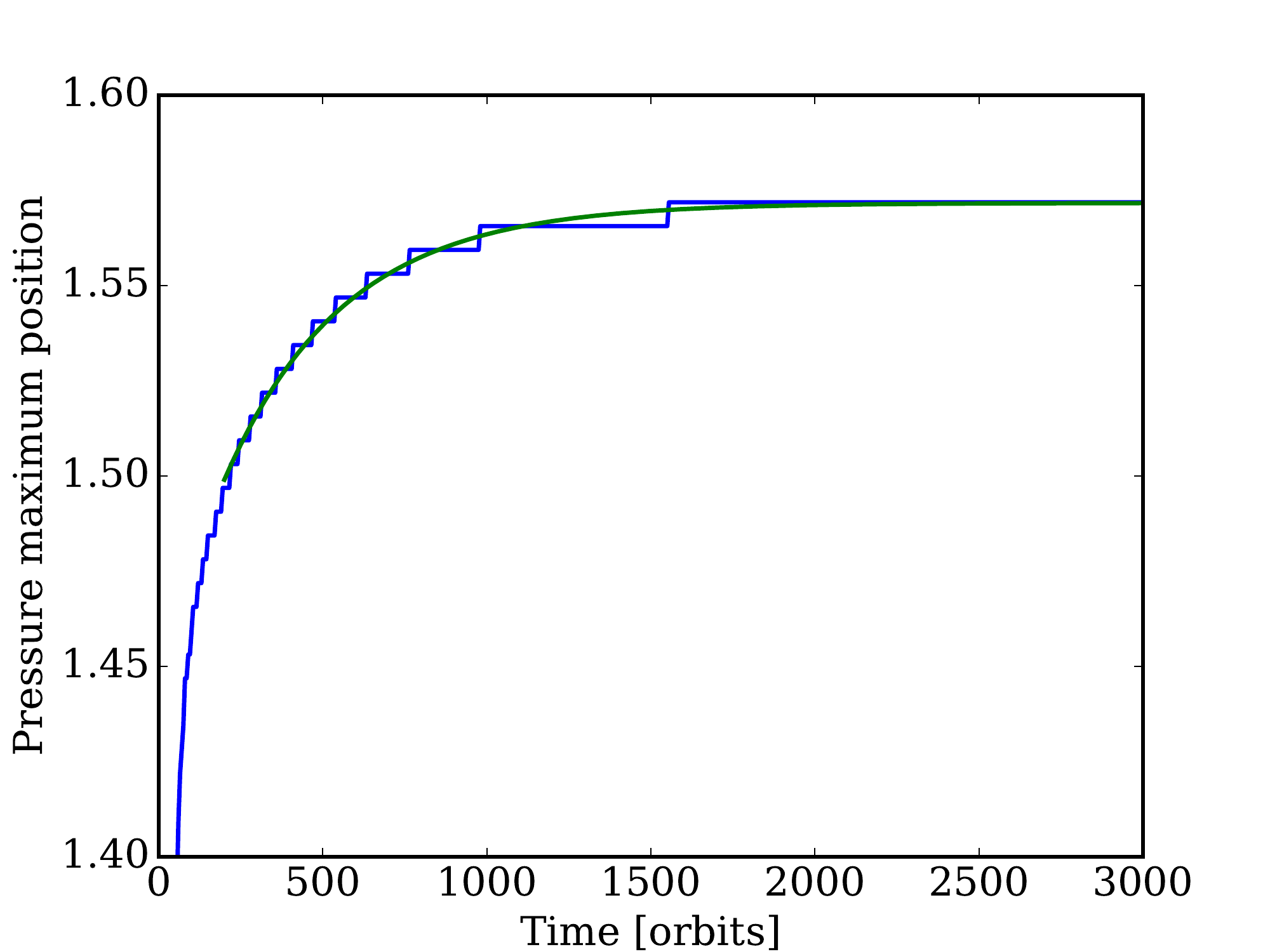}
\caption{Evolution with time of the location of the pressure maximum outside the planet orbit (blue line) for the reference case of $480 M_\oplus$ in the ``hot'' disc. We also fit the data with an exponential (green line) to show that convergence has been reached.}
\label{fig:press_max_t}
\end{figure}

\begin{figure}
\includegraphics[width=\columnwidth]{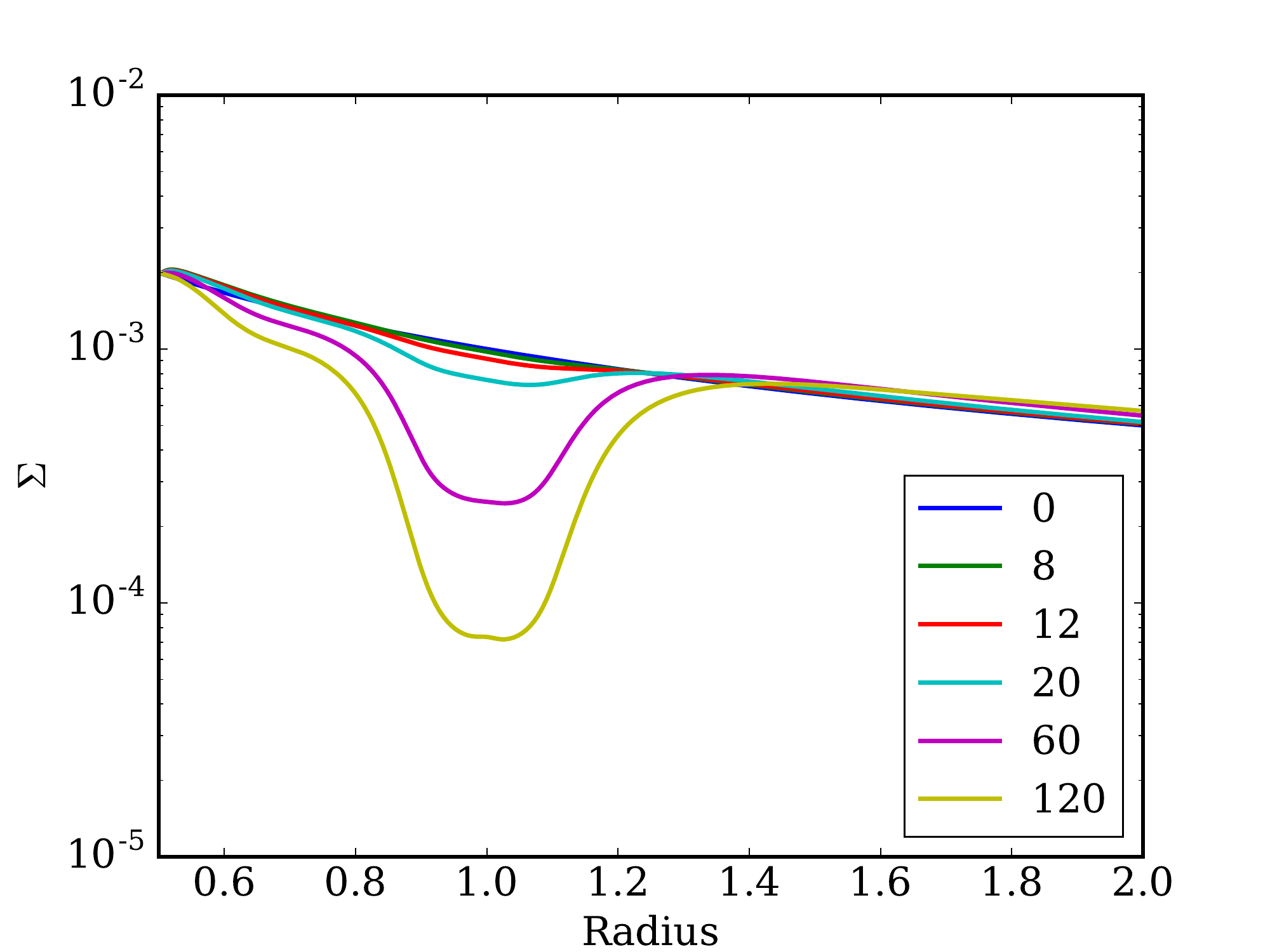}
\caption{Azimuthally averaged gas surface density for the fiducial model for the different planet masses.}
\label{fig:gas_1d}
\end{figure}

We find that it is necessary to integrate for a relatively long time to reach a consistent steady state. We show the time evolution of the surface density for a representative case of 60 $M_\oplus$ in figure \ref{fig:gas_t}. Different lines correspond to different snapshots from the simulations: 150 orbits (blue dotted line), 400 orbits (green dotted-dashed line), 1000 orbits (red dashed line) and finally 3000 orbits (solid cyan line). While the gas converges reasonably close to the final result on a timescale of $\simeq 100$ planetary orbits, the depth of the gap still slowly evolves over a timescale of $\simeq 10^3$ planetary orbits, over which the depth typically changes of a factor $<2$. This result is consistent with what other authors have found previously when looking at the timescale of gap opening\citep{2004ApJ...612.1152V,2013ApJ...769...41D,2014ApJ...782...88} and can be understood in terms of the timescale to reach gap opening being roughly the viscous timescale across the gap. For reference, if we assume that the perturbed region has a width $\Delta$ of few scale-heights: $\Delta = x H$ (with $x$ a constant with a value of a few), we find that the timescale for gap opening is $x^2/(2\pi \alpha) \sim 150 x^2 $ orbits. { \citet{CridaOpeningCriterion} found that instead the relative surface density profile, i.e. $1/\Sigma \, \mathrm{d}\Sigma/\mathrm{d}r$, is established on a much faster timescale}. While the gas evolution has few observational implications in itself, we find that the dust can change more significantly over this time-scale. In particular, the dust is sensitive to the pressure profile of the gas. If the planet is massive enough, it creates a pressure maximum in the gas that will trap dust \citep{Rice2006}, as we detail in section \ref{sec:dust_vr}. The location of this maximum has therefore a deep impact on the dust structure and we find that it reaches convergence after the stated time interval. We show in Figure\,\ref{fig:press_max_t} the evolution with time of the location of the pressure maximum for { an illustrative case of a planet of $480 M_\oplus$ in the ``hot'' disc, that is, $H/R=0.1$. We choose this disc because the viscous time-scale in this disc is faster than for the other aspect ratios, which allows us to explore more the evolution in time of the pressure maximum. The curve (blue line) can be fitted with an exponential (green line) with a tapering timescale of $\sim 370$ orbits; the value that the exponential tends to is almost indistinguishable from the value after 3000 orbits. For the fiducial model, an exponential convergence is not so clear and the curve can be fitted equally well by a logarithm. Assuming that we can rescale the time scale with $(h/r)^2$, we can estimate the error in the
location of the pressure maximum in the fiducial model that is introduced by only running for 3000 orbits to be 0.2 H, by estimating the error for the ``hot'' case if we had stopped after 750 orbits. Even for the ``cold'' case, the error is less than $H$. We remark that current observational facilities are not able to resolve a scale height in discs at a distance of 30 AU from the star, so that our estimates are accurate enough for observational applications. In what follows we choose then to present our results after 3000 orbits.}

Figure \ref{fig:gas_1d} shows the azimuthally averaged surface density of the gas for the fiducial model after 3000 planetary orbits. In figure \ref{fig:gas_1d} it can be seen that there is little modification to the gas surface density for the least massive cases. The $120 M_\oplus$ planet produces a gap that involves a reduction in surface density by a factor 10; the $60 M_\oplus$ one produces a partial gap that involves a reduction by a factor 4. The relatively small depth of these gaps is not surprising given that for the planet masses considered here only the most massive case is in the gap opening regime. Two commonly used \citep{LinPapaloizou93} criteria for gap opening are the so-called thermal criterion, according to which the planet Hill radius $R_H$ must be greater than the vertical scale height of the disc $H$:
\begin{equation}
q \gtrsim 3 \left( \frac{H}{R} \right)^3,
\label{eq:thermal}
\end{equation}
where $q=M_p/M_\ast$ (we assume $M_\ast = 1 M_\odot$ for simplicity in what follows), and the viscous criterion, according to which the time-scale to open the gap must be smaller than the time-scale for viscosity to close it:
\begin{equation}
q \gtrsim \left( \frac{27\pi}{8}\right)^{1/2} \left(\frac{H}{R} \right)^{5/2} \alpha^{1/2}.
\end{equation}
Substituting numerical values for our case gives a threshold mass ratio of $4 \times 10^{-4} \simeq 120 M_\oplus$ for the thermal criterion and $5.5 \times 10^{-5} \simeq 16 M_\oplus$ for the viscous criterion. This means that for our choice of parameters the pressure forces are the dominant ones trying to close the gap and therefore only the thermal criterion should be considered. To be in the regime where also viscosity is important would require to consider a lower value of $H/R$ or a bigger value of $\alpha$. Note that however, due to the shallow dependence on $\alpha$ and the similar dependence on $H/R$, in practice this happens only for quite extreme values of the parameters. \citet{CridaOpeningCriterion} collected the two criteria in only one that accounts for both conditions:
\begin{equation}
\frac{3}{4} \frac{H}{R_H} + \frac{50 \nu}{q r^2_p \Omega_p} \lesssim 1.
\end{equation}
{ The interested reader can consult \citealt{2014prpl.conf..667B} for an alternative formulation of the same criterion, that gives explicitly the mass ratio $q$}. The numerical factors in this equation are slightly different so that the threshold mass ratio is approximately $1.5 \times 10^{-4} \simeq 45 M_\oplus$. However note that their definition of gap is a factor of 10 reduction in the surface density, which actually happens in our simulations only for the most massive planet. Simplified criteria like the ones we quote here can be used only as order of magnitude estimates and one should not overinterpret them. What matters for the purpose of this paper is that these planets are not in a regime where they open deep gaps in the gas surface density such those opened by Jupiter mass planets. Finally, note the very steep dependence of the thermal criterion on the disc temperature, which will be important in the rest of the paper.


\begin{figure*}
\includegraphics[width=\textwidth]{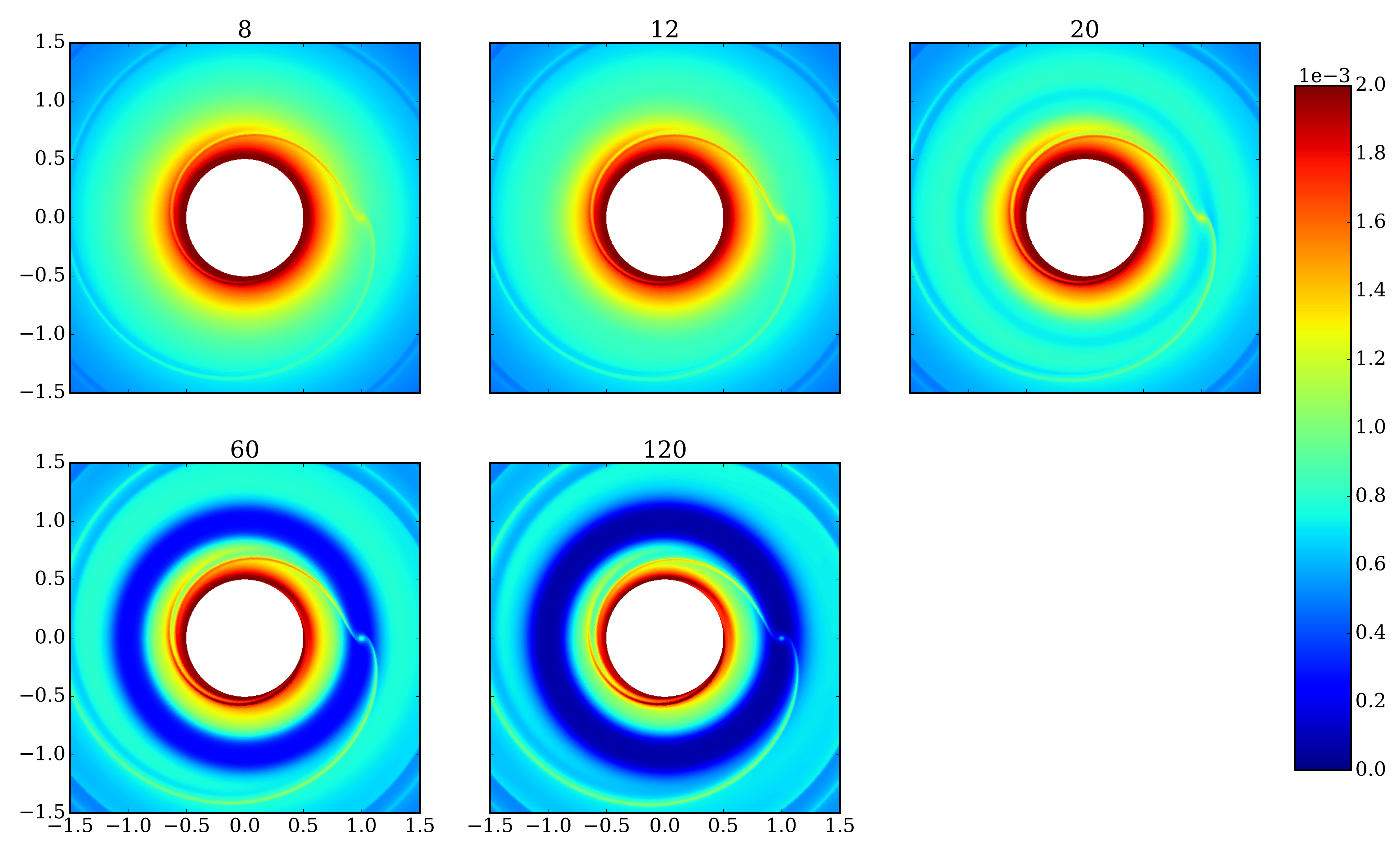}
\caption{Gas surface density for the different planet masses (as indicated in Earth masses).}
\label{fig:gas_2d}
\end{figure*}

Figure \ref{fig:gas_2d} shows the 2d gas surface density. Note that although the gap starts to be visible only for the $60 M_\oplus$ case, all planets generate a clear spiral in the 2d gas surface density. Such spirals however are too thin and with a contrast that is too low to be observed. We will show this in the next section where we show simulated observations. Even in the case of very massive planets, the conclusions of \citet{AttilaSpirals} were that it is extremely hard to detect spirals such as the ones recently imaged in the scattered light \citep[e.g.,][]{2012ApJ...748L..22M,Garufi2013,2015ApJ...813L...2W,2015A&A...578L...6B}. It should therefore not be surprising that spirals are not detectable for these low mass planets.

\subsection{Dust surface density}
\label{sec:dustsigma}

\begin{figure*}
\includegraphics[width=\textwidth]{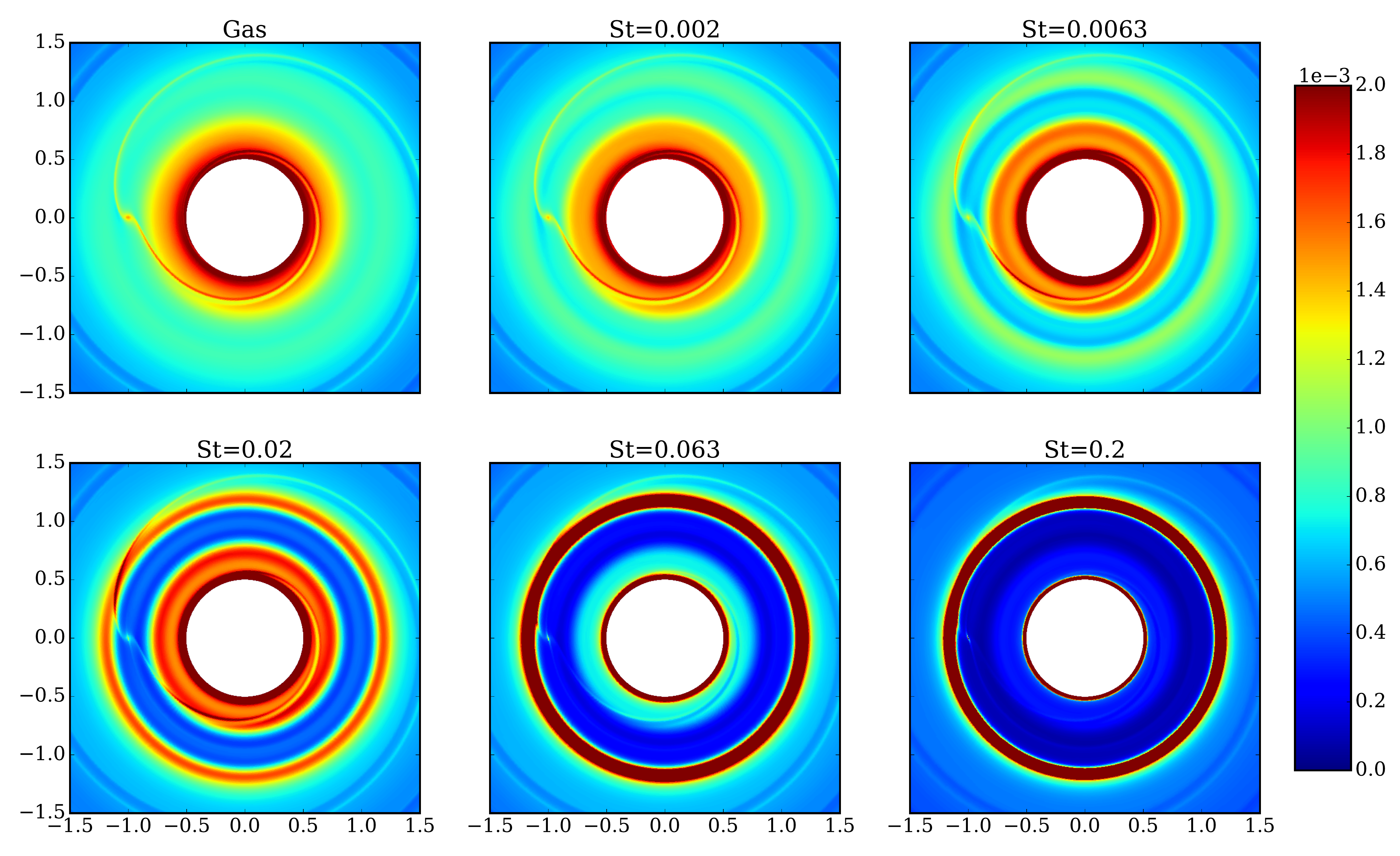}
\caption{Dust surface density for the $20 M_\oplus$ planet. Different panels correspond to different Stokes numbers as shown in the legend. For reference we show also the gas surface density.}
\label{fig:dust_20earth_2d}
\end{figure*}

\begin{figure}
\includegraphics[width=\columnwidth]{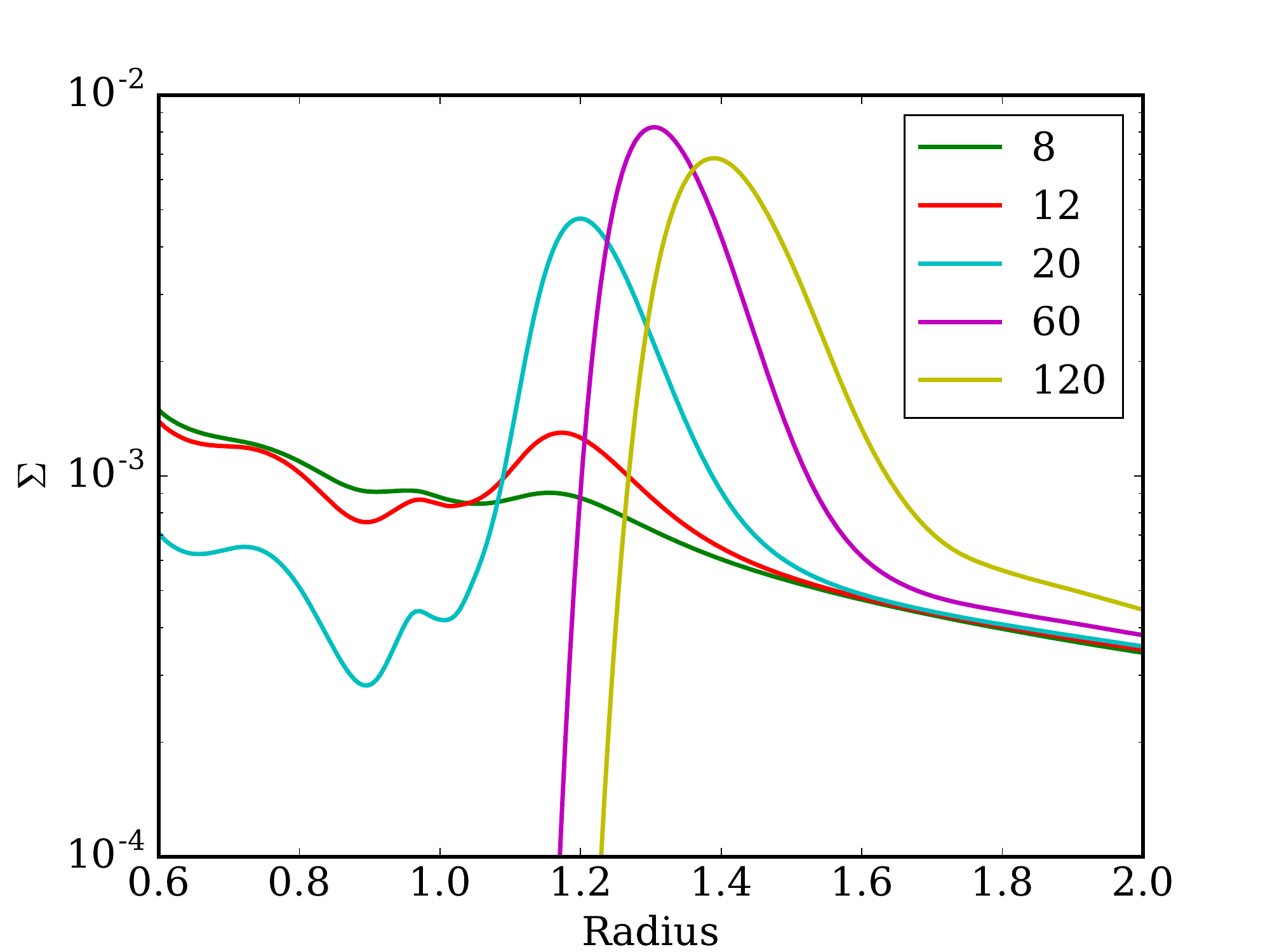}
\caption{Dust surface density with $St=0.02$ for the different planet masses.}
\label{fig:dust0.02}
\end{figure}

\begin{figure}
\includegraphics[width=\columnwidth]{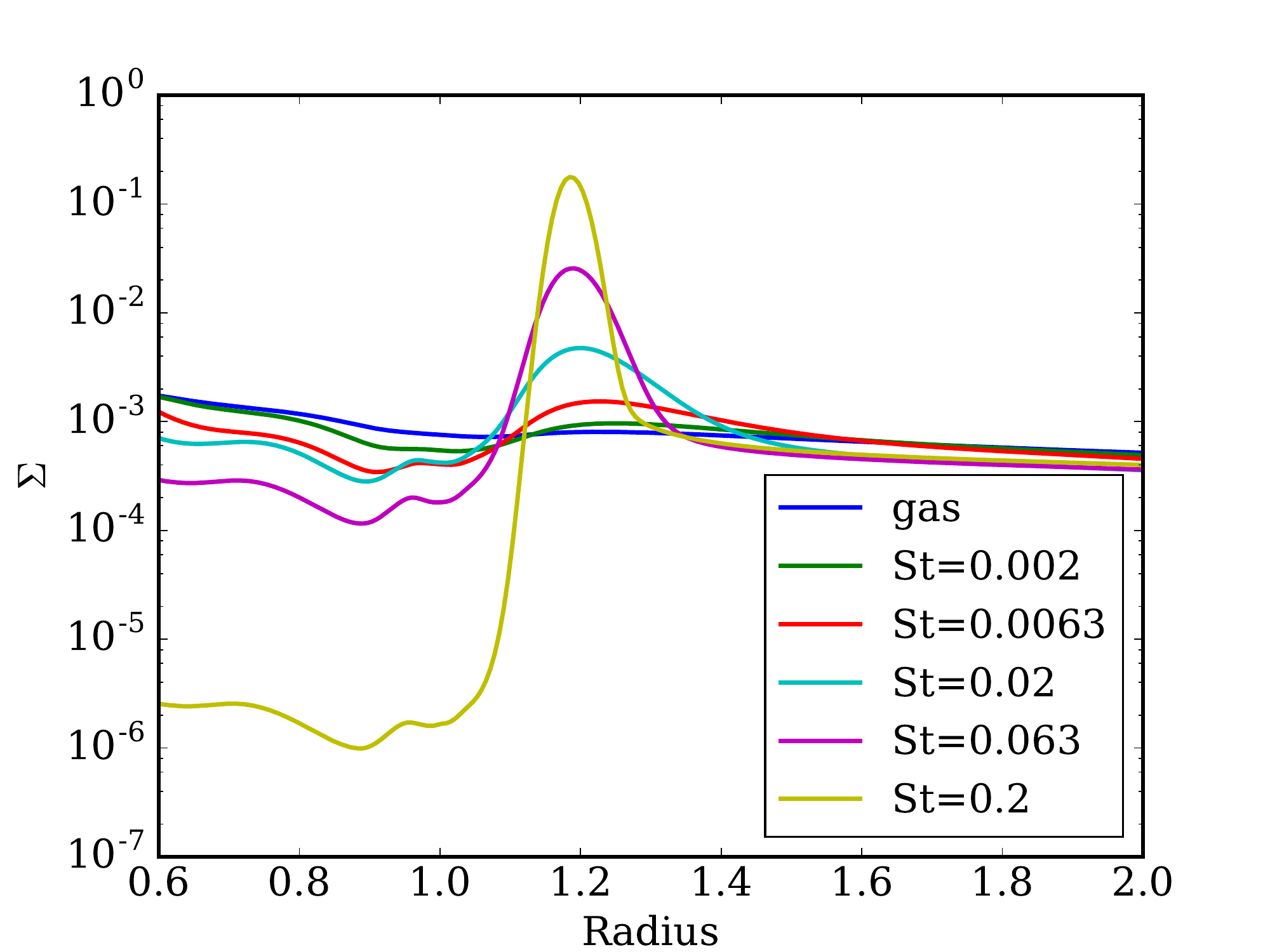}
\caption{Dust surface density for the $20 M_\oplus$ planet. Different lines correspond to different Stokes numbers as shown in the legend. For reference we show also the gas surface density.}
\label{fig:dust_20earth}
\end{figure}

We begin by reporting the features produced in the dust by planets of varying mass; the observability of these features will be discussed in section \ref{sec:sim_obs}. It is impossible to show all the results from our simulations in the limited space of this paper. Instead, it is more instructive to understand the trends with planet mass and particle size and to determine what drives this behaviour. As a representative example, we show for a 20 $M_\oplus$ planet how the dust surface density varies with Stokes number (figure \ref{fig:dust_20earth_2d})\footnote{Note that the multi-fluid simulations are not sensitive to the normalisation of the dust surface density; for illustrative purposes in figures \ref{fig:dust_20earth_2d}, \ref{fig:dust0.02} and \ref{fig:dust_20earth} we have normalised each dust surface density so that it has initially the same value as the gas surface density.}. For sufficiently small dust particles, which are well coupled to the gas, the depth of the gap in the dust density should be the same as the gas. For our smallest size we indeed see only slight difference between dust and the gas; however, it is possible to see that the dust density at the edge of the gap is slightly larger. By the next size we simulate ($St = 0.063$) there is already an order unity variation in the surface density, which continues to increase with increasing dust size. 

From these images it is clear that the structures in the dust density are largely azimuthally symmetric (for the planet masses we consider here; for massive enough planets this is not necessarily true), with the exception of the spiral feature, which is thin and only has a small density enhancement associated with it. Thus it is instructive to consider the trends  seen in azimuthally averaged profiles. From the azimuthally averaged profiles (figures \ref{fig:dust_20earth} \& \ref{fig:dust0.02}) we see that both increasing the planet mass and particle size leads to stronger effects in the dust. The trends with planet mass for a fixed Stokes number (here $St=0.02$, figure \ref{fig:dust0.02}) show that the depth of the gap depends clearly on the planet mass, with more massive planets creating deeper gaps. The 8 $M_\oplus$ planet does not significantly modify the dust surface density (the variation is only a few percent even for the largest dust size we simulate), while the 12 $M_\oplus$ produces variations of order unity. The 60 and 120 $M_\oplus$ planets actually create a \textit{hole} rather than a narrow \textit{gap}, meaning that the surface density of the dust inside $\sim 1.2$ times the orbital radius of the planet is severely depleted of dust; the 20 $M_\oplus$ planet is in between a gap and a hole. There is thus only a limited range of planet masses that create detectable gaps for this Stokes number; planets that are massive enough will eventually form holes. Similar trends hold with particle size, with the 20 $M_\oplus$ planet opening a gap in the $St = 0.2$ dust. One notable fact is that while the location of the peak dust density in the ring depends on planet mass, it is not a function of grain size.

Even for planets that just open gaps we find that the gaps opened in dust can be much deeper than the ones in the gas, consistent with what was found by previous authors \citep{PaardekooperMellema2006,zhu2014}.  Compare for example the $20 M_\oplus$ case {in figure \ref{fig:dust0.02}} with Figure\,\ref{fig:gas_1d}, where the presence of a planet is only barely visible by looking at the surface density. In addition, the more massive cases we simulate not only show a gap or a hole, but also a bright ring in the surface density outside the orbital location of the planet. The enhancement of this peak, and even more interestingly its location, depend on the planet mass ({figure \ref{fig:dust0.02}}). We will investigate the reasons for this morphology in section \ref{sec:dust_vr}. We do not find any feature like the one that \citet{PaardekooperMellema2006} found at outer resonances, in line with the results of \citet{zhu2014}.

\subsection{Dust radial velocity}
\label{sec:dust_vr}

Since the dust density is azimuthally symmetric, this suggests that the dust dynamics and resulting surface density can be understood in terms of a one dimensional model in which the continuity equation is just
\begin{equation}
\frac{\partial \Sigma_d}{\partial t} + \frac{\partial (R \Sigma_d v_{d,R})}{\partial R}  =0.
\label{eq:steady_state}
\end{equation}
This equation can be closed with the knowledge of dust velocity; for particles with $St \ll 1$ we can make use of the short-friction time approximation in which the dust velocity is given by
\begin{equation}
v_d = v_g + t_s \frac{\nabla P}{\rho} + v_{\rm D},
\label{eq:sft}
\end{equation}
where the last term takes into account the diffusive flux. The expression provides a simple way of dissecting the dust velocity into the contribution from the gas and that due to the pressure gradient, since apart from at the gap edge $v_{\rm D}$ is generally small. Thus the pressure term is responsible for the differences between the dust and the gas velocities; in turn, these affect the surface densities. It is straightforward to realise looking at the expression that the importance of the pressure gradient increases with the particle size. 

\begin{figure*}
\includegraphics[width=\columnwidth]{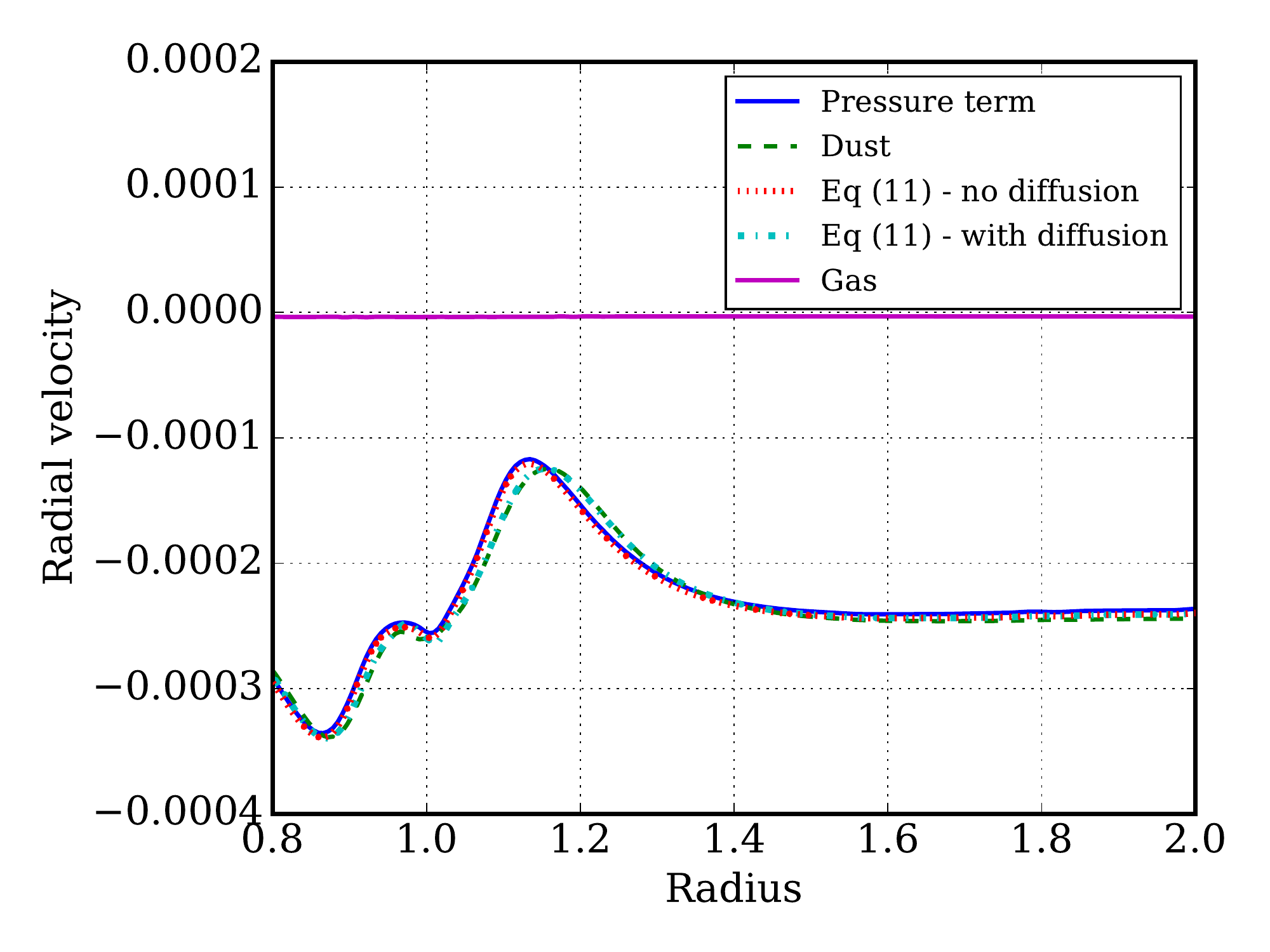}
\includegraphics[width=\columnwidth]{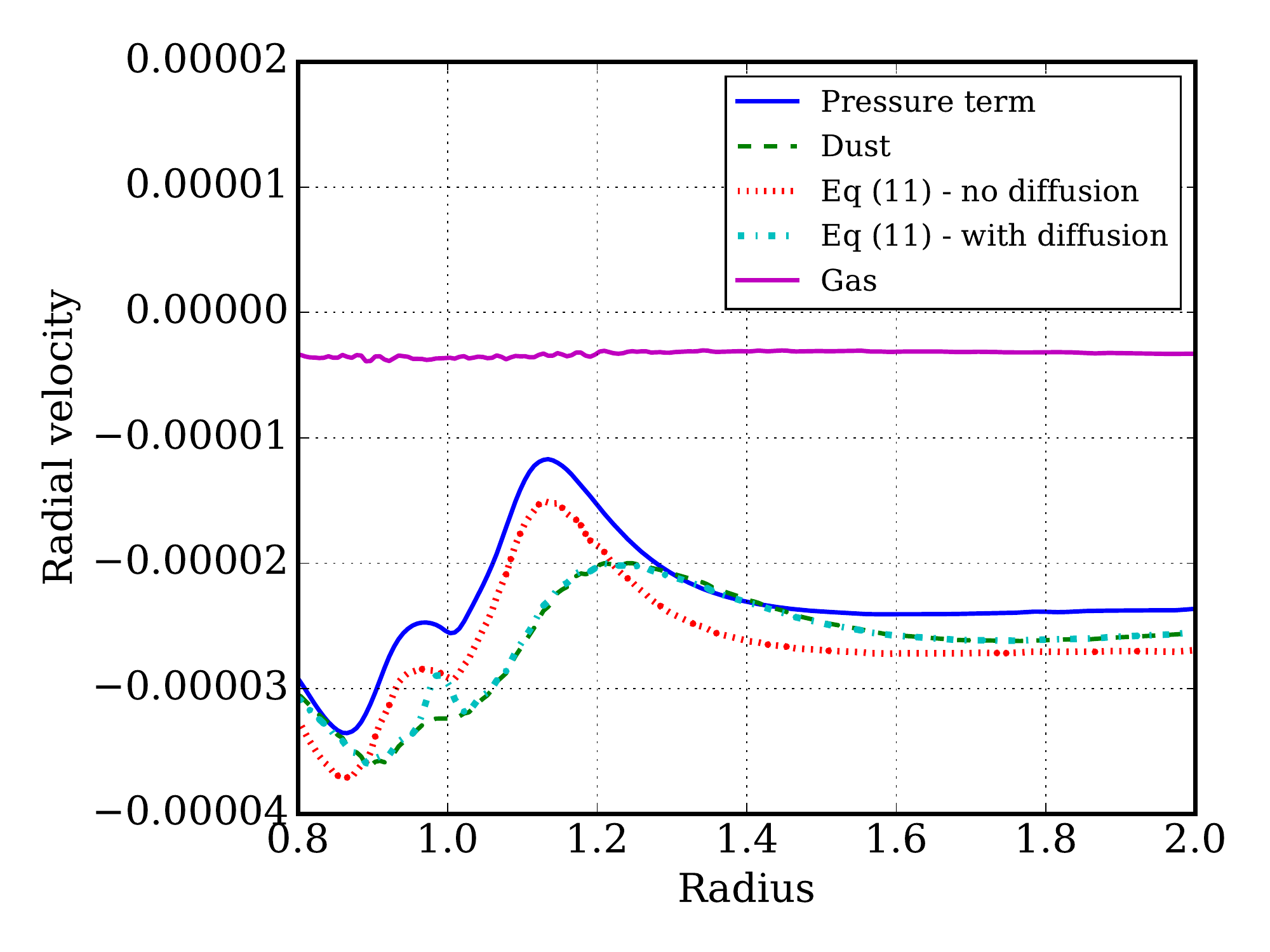}
\caption{Dissection of the dust radial velocity for two sizes for the $12 M_\oplus$ planet. The plots shows the pressure gradient term from equation (\ref{eq:sft}) (solid blue line), the actual value of the velocity from the simulation (dashed green line), the estimate of equation (\ref{eq:sft}) with (dotted-dashed cyan line) and without (dotted red line) the diffusion term and finally the gas radial velocity from the simulation (solid purple line). Left panel: $St=0.07$. Right panel: $St=0.007$.}
\label{fig:vrad_dissec}
\end{figure*}

Figure \ref{fig:vrad_dissec} shows a comparison of the actual dust radial velocity from the simulation, azimuthally averaged, with the approximation given by equation (\ref{eq:sft}) (where again quantities have been azimuthally averaged) for two dust sizes for the $12 M_\oplus$ planet (left panel: $St=0.07$; right panel: $St=0.007$). Specifically, the plots shows the pressure gradient term from equation (\ref{eq:sft}) (solid blue line), the actual value of the velocity from the simulation (dashed green line), the estimate of equation (\ref{eq:sft}) including (dotted-dashed cyan line) or not (dotted red line) the dust diffusion term (that is, $v_{\rm D}$) and finally the gas radial velocity from the simulation (solid purple line). We can see that the pressure gradient is always negative. To state it in another way, the planet is not massive enough to create a pressure maximum outside the gap edge. Since this condition is required to trap dust particles (because it provides a location where the dust radial velocity is zero), the planet is unable to trap dust and open up a \emph{hole} \citep{Rice2006}. However, there is still a region where the pressure gradient is weaker, resulting in a smaller radial velocity. This is then a ``traffic jam'' rather than a dust trap, in which a higher dust density region is associated with the lower velocity. We conclude then that the existence of a pressure maximum is not a necessary condition to affect the dust surface density. { While this paper was being refereed, \citet{2016arXiv160207457D} proposed another mechanism that can open gaps in the dust without requiring a pressure maximum. We note that their mechanism is different from the one we describe here and applies to grains with a large Stokes number (they assume $St=10$).} 

We now illustrate that the dust is in (or close to) steady-state. By setting $\partial \Sigma_d / \partial t = 0$, the continuity equation reduces to $\Sigma \propto 1/(R v_R)$, from which the constant of proportionality can be fixed using the surface density far from the planet.
\begin{figure}
\includegraphics[width=\columnwidth]{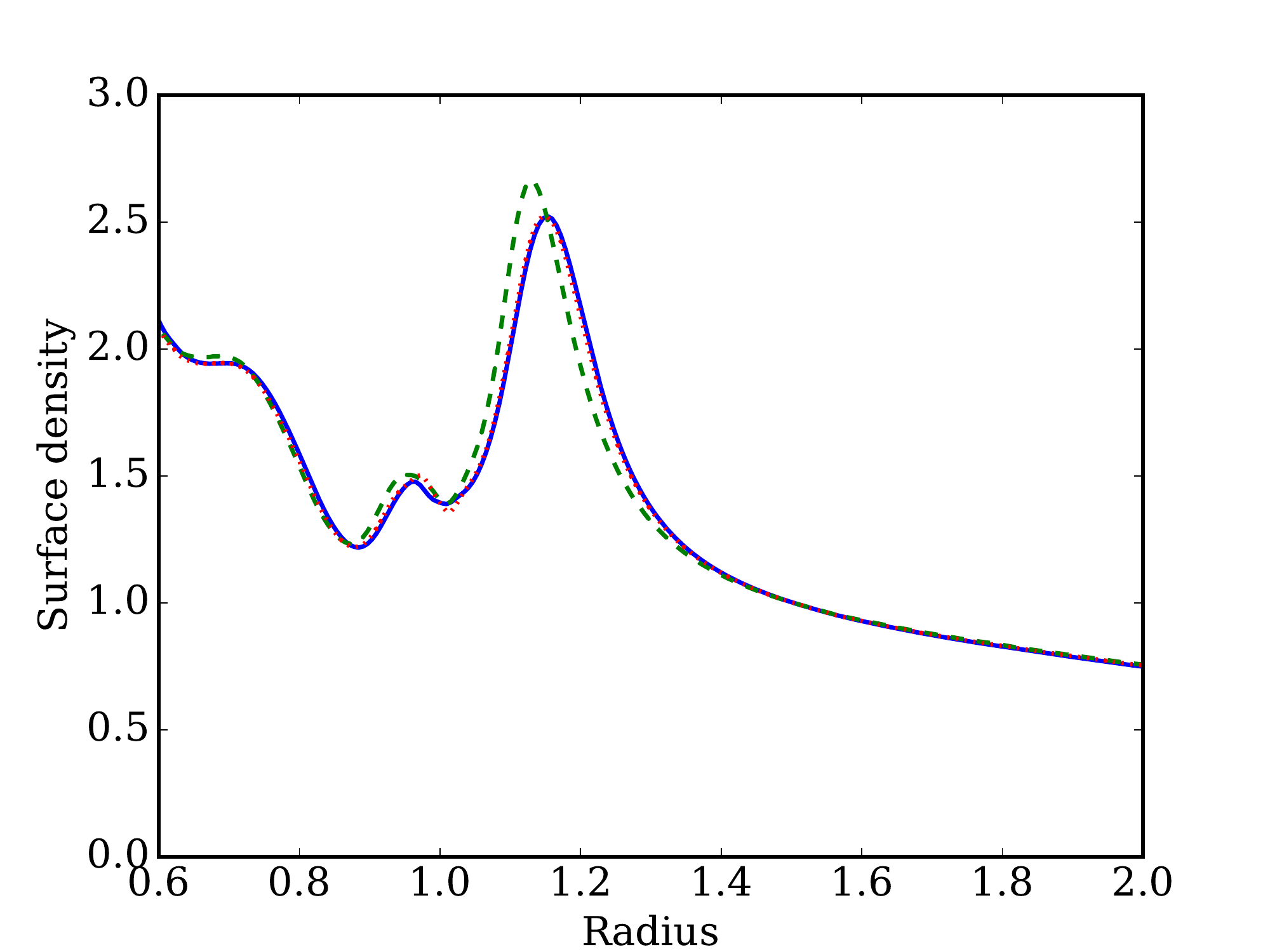}
\caption{Comparison between the steady state surface density predicted by equation (\eqref{eq:steady_state}) (green dashed line: without including dust diffusion; dotted red line: including dust diffusion) and the actual value from the simulation (solid blue line) for the 12 $M_\oplus$ planet, $St=0.07$. }
\label{fig:steady}
\end{figure}
We demonstrate this in Figure \ref{fig:steady}, which shows a comparison between the steady state expected from the continuity equation and the actual surface density from the simulation (solid blue line). We show two different steady state predictions; the difference between them is that the dashed green line neglects dust diffusion (i.e., it is computed using the dotted red line of figure \ref{fig:vrad_dissec}) whereas the dotted red line includes it (i.e., it is computed using the dotted-dashed cyan line). The remarkable agreement confirms that the structure that we see in the surface density is sculpted by the structure in the velocity, namely by (mostly) the pressure gradient (with a minor contribution from diffusion). This also means that in this particular case we could potentially have deduced the dust structure (neglecting dust diffusion) purely from the gas surface density and radial velocity, without running any multi-fluid simulation. However, for the right panel of figure \ref{fig:vrad_dissec} ($St=0.007$) dust diffusion is very important in smoothing out the structures carved by the pressure gradient. In this case, since the dust diffusion depends on the dust density, multi-fluid simulations are needed. 

\begin{figure}
\includegraphics[width=\columnwidth]{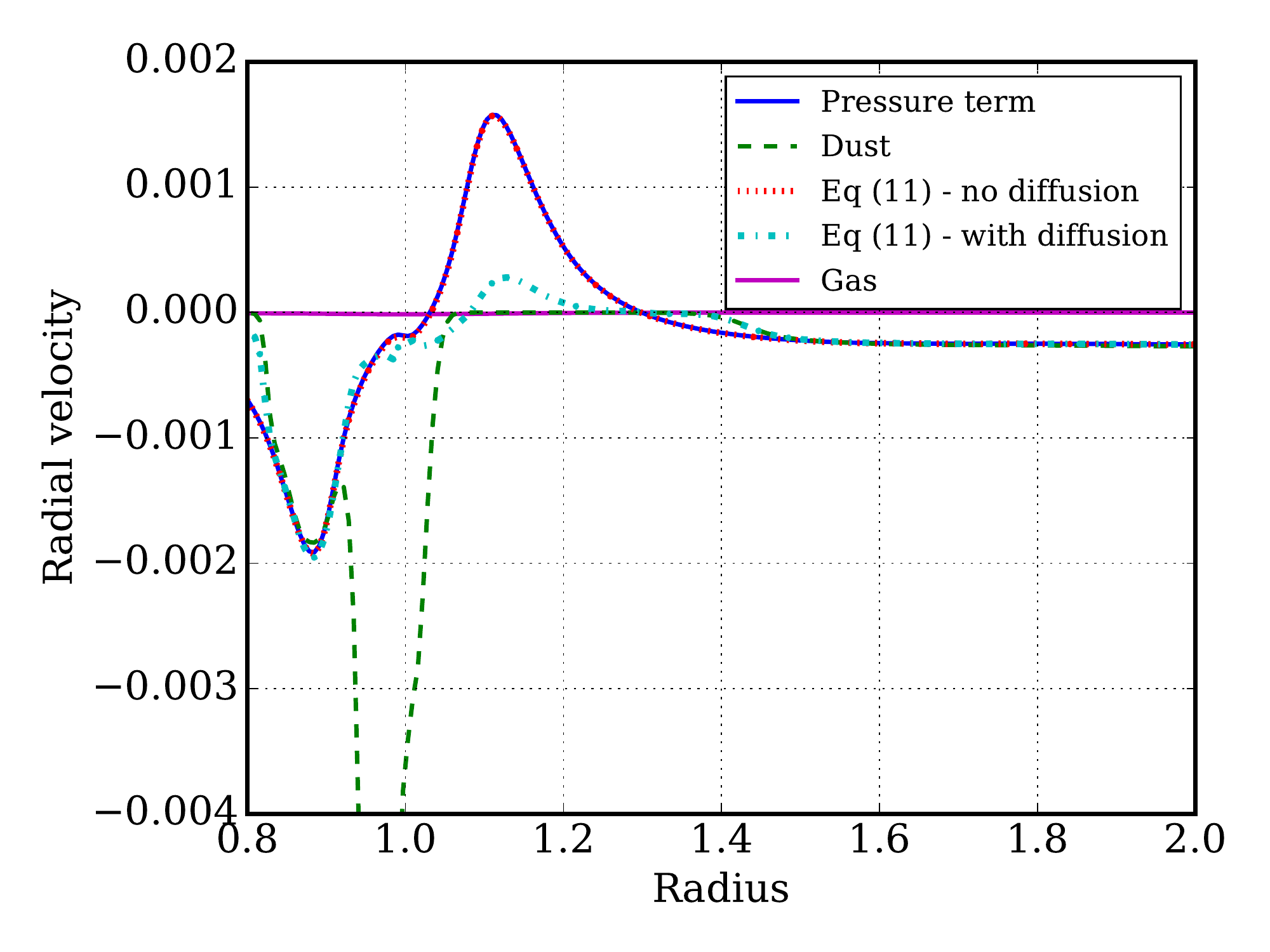}
\caption{Dissection of the dust radial velocity for the 60 $M_\oplus$ planet (see Figure \ref{fig:vrad_dissec} for the meaning of the different lines), $St=0.07$. }
\label{fig:vrad_dissec60}
\end{figure}

For more massive planets, the planet creates a maximum in the pressure and we have an actual pressure trap for large enough particles. We show in figure \ref{fig:vrad_dissec60} one such case with a 60 $M_\oplus$ planet\footnote{The 20 $M_\oplus$ planet also creates a pressure maximum, although only very shallow} and $St=0.07$ dust, where the meaning of the lines is the same as in Figure\,\ref{fig:vrad_dissec}. For this particular size the pressure gradient term is much larger than the gas velocity; however the $t_s$ dependence of this term (equation \ref{eq:sft}) means that there is a minimum dust size that can be trapped, which for a $60M_\oplus$ planet requires $St \lesssim 3 \times 10^{-4}$. Particles above this size can only break through this trap via diffusion; thus, as noted by previous authors \citep{Rice2006,Zhu2012,Pinilla12}, dust will continue to pile up at the location where the pressure gradient vanishes and a maximum in the dust surface density will occur. We will comment more on the location of the maximum in the dust in section \ref{sec:measure_mass}. Diffusion smooths out the maximum as it tends to cancel gradients in the dust concentration, so that the surface density will not grow indefinitely at the location of the pressure maximum. However, this takes a long time as reaching a steady state requires the diffusive velocity to be comparable to the pressure gradient term, and therefore a very big dust accumulation, which is provided by dust drifting from the outer part of the disc. Indeed, in our simulation we see that, for planets that create traps, the surface density at the pressure maximum is still increasing even after 3000 orbits, and no steady state has been reached yet (differently from the ``traffic jam'' case shown previously). As shown in the previous section, the trap cuts out most of the supply to the inner disc, so that the planet opens a hole rather than a gap for this particular case (which for { dust} sizes close to the minimum hole opening size can be partially filled by diffusion). Figure \ref{fig:vrad_dissec60} also shows how the estimate provided by equation (\ref{eq:sft}) breaks down close to the planet. Our interpretation is that 2d effects in this region cannot be neglected. In this case only 2d dust dynamics simulations can recover the correct result.

We have repeated the same analysis shown here also for the ``cold'' and ``hot'' disc and confirm that a pressure maximum appears for the same ratio of the planet mass to thermal gap opening mass. { To summarise, in our simulations} the existence of a pressure maximum happens at a mass $\gtrsim 0.2 M_{th} \sim 20 ((H/R)/0.05)^3 M_\oplus$, where $M_{th}$ is gap opening mass given by the thermal criterion. { We then confirm the results of \citet{2014A&A...572A..35L}, who in a different context found roughly the same value through 3D simulations. The fact that 2D and 3D simulations give a very similar result is very encouraging and means that this result can be taken to be robust.}

\section{Simulated observations}
\label{sec:sim_obs}


To study the observability of the gaps we also calculate simulated observations at NIR, MIR and sub-millimetre wavelengths. In the NIR  we study the observability of the gaps with the current state-of-the-art imaging instrument, SPHERE on VLT \citep{2008SPIE.7014E..18B}. While the resolution of the currently available MIR instruments is too low to allow the detection of planets in this mass regime, this will change in a few years with the advent of 30m class telescopes. We thus compute also MIR images, with the resolution of the planned NIR/MIR instrument, Mid-infrared E-ELT Imager and Spectrograph (METIS)  on the E-ELT,  to assess the potential of this wavelength in detecting low mass planets. Since both of these instruments provide diffraction-limited resolution due to the extreme adaptive optics used, we simulate the observations by convolving the resulting images with a 2D Gaussian kernel. The full-width at half maximum (FWHM) was taken to be $\lambda/D$, where $\lambda$ is the wavelength of observation and $D$ is the telescope diameter.


In case of the  sub-millimetre images we simulate images using the Common Astronomy Software Applications\footnote{http://casa.nrao.edu/index.shtml} (CASA) v4.2.2. We use the {\tt simobserve} task to simulate the observed visibilities then the images are calculated from the visibilities using the {\tt clean} task. The full 12-m array is used in two different configurations resulting in 0.027\arcsec and 0.091\arcsec resolution, respectively, at 880\,$\mu$m. The source declination is taken to be $\delta=-25^\circ$. We simulate observations with 1\,h integration time using the full 7.5\,GHz bandwidth and 0.913\,mm precipitable water vapour, typical for Band\,7 observations. For all simulated observations we assume that the distance to the source is 140\,pc.

\subsection{Fiducial model}

\begin{figure*}
\includegraphics[width=\textwidth]{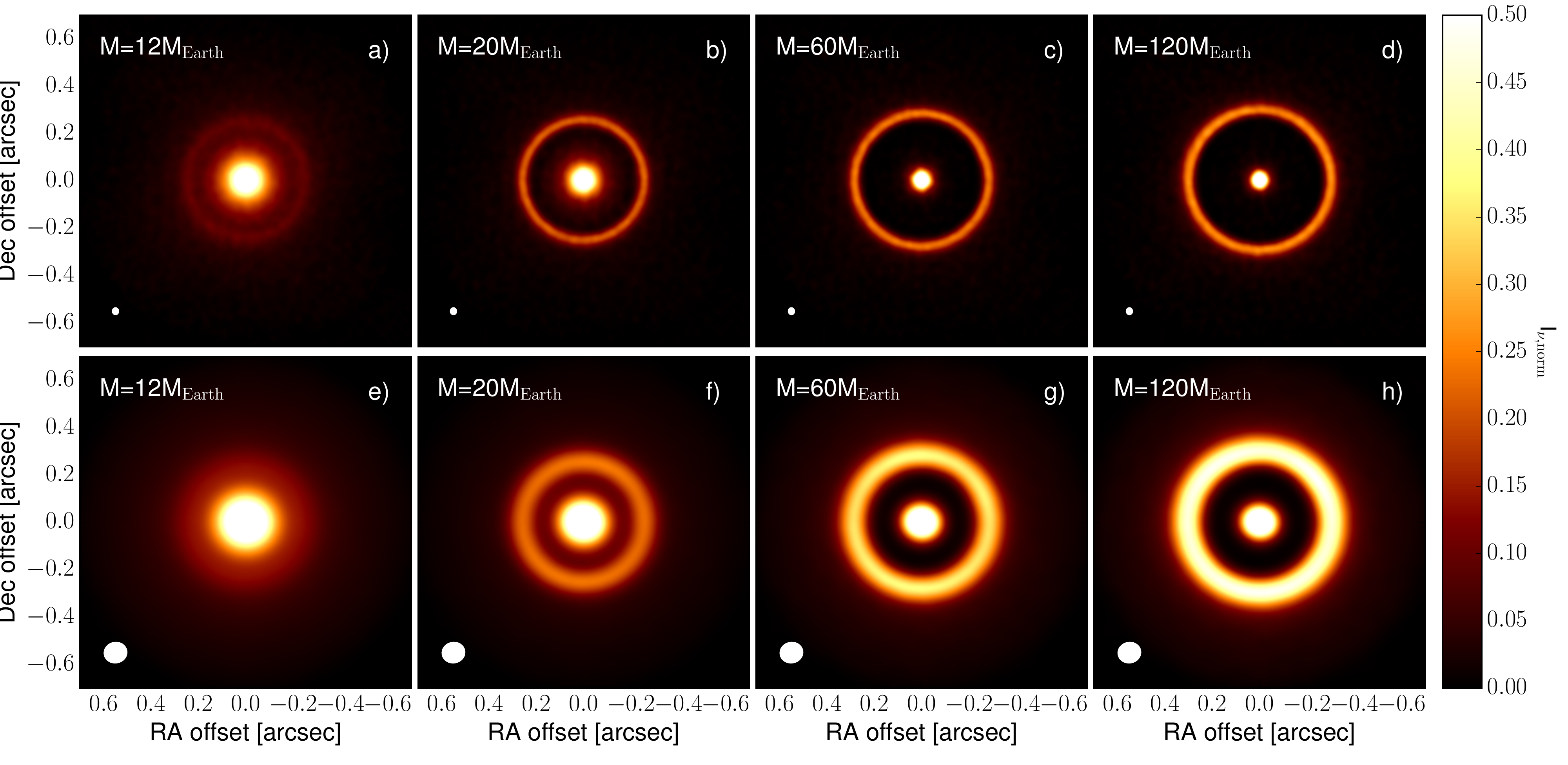}
\caption{Sub-mm images for four different planet masses for the fiducial model. We show two different resolutions of 0.025" (top row) and 0.091" (bottom row). The size of the synthesised beam is shown as a filled white ellipse in the bottom left corner of each panel.}
\label{fig:alma}
\end{figure*}

\begin{figure*}
\includegraphics[width=\textwidth]{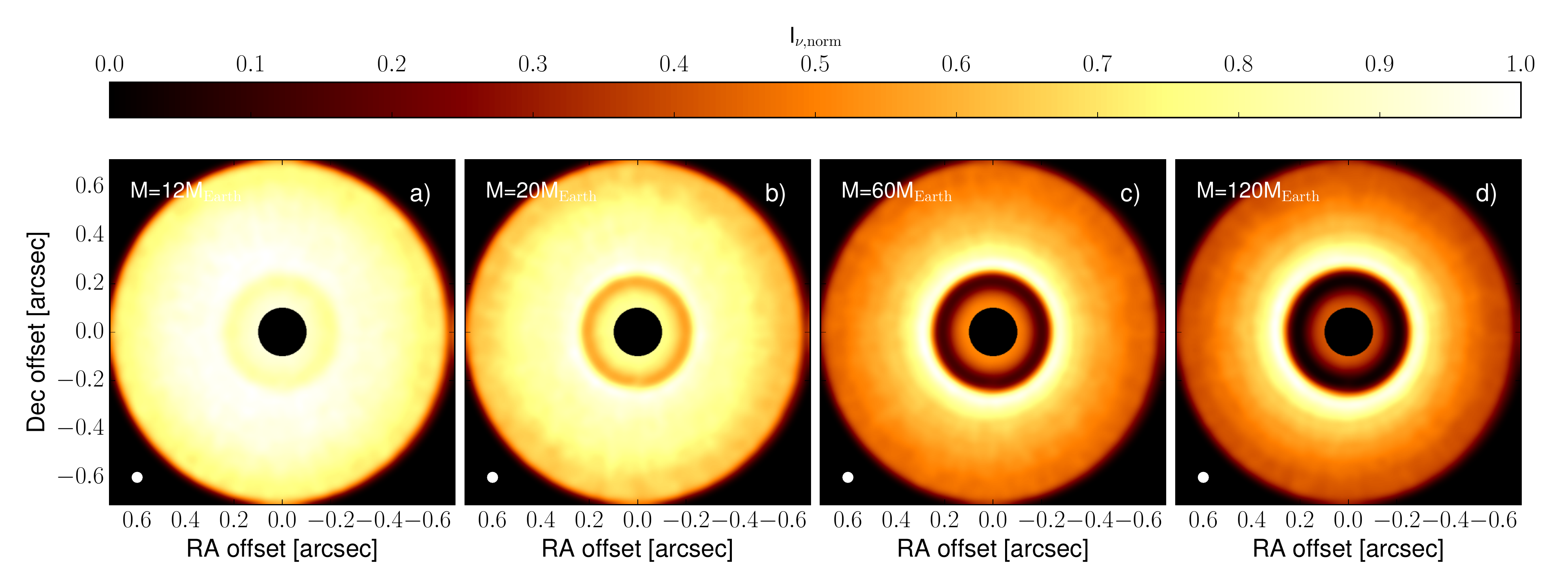}
\caption{Images in the scattered light for the same planet masses as the previous figure. The resolution of the radiative transfer images was degraded to that of SPHERE on VLT (0.04\arcsec), by convolving the images with a 2D Gaussian kernel. The central 0.1\arcsec of the disc was masked to
simulate the effect of a coronograph. The images are scaled with the square of the radial distance from the central star.  The size of the PSF is shown 
in the bottom left corner as a filled white ellipse.}
\label{fig:nir}
\end{figure*}

\begin{figure*}
\includegraphics[width=\textwidth]{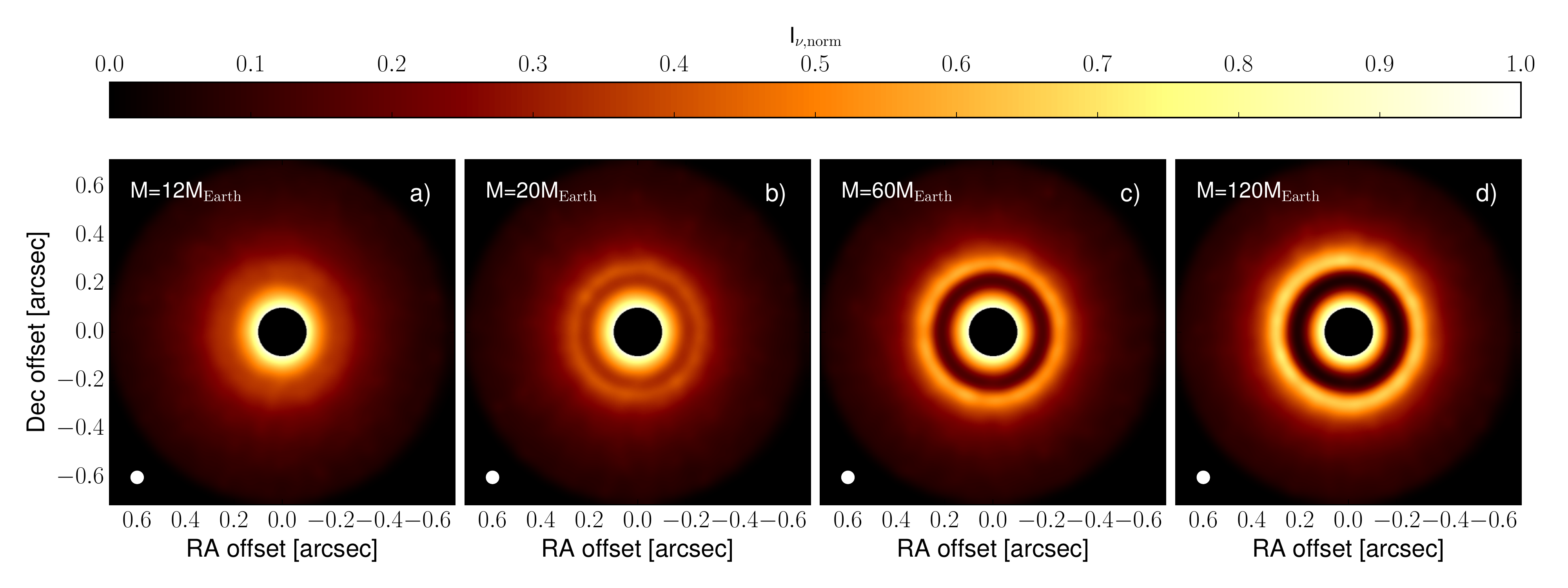}
\caption{Images in the MIR of the thermal emission from the disc  for the same planet masses as in figure \ref{fig:alma}. The resolution of the radiative transfer images was degraded to that of the future METIS instrument on the European ELT by convolving the images with a 2D Gaussian with 
an FWHM of 0.05\arcsec. The images are scaled with the square of the radial distance from the  star and the central 0.1\arcsec of the disc was masked to simulate the effect of a coronograph. The size of the PSF is shown in the bottom left corner as a filled white ellipse.
Current instruments lack the spatial resolution to be able to resolve the morphologies we describe in this paper.
}
\label{fig:metis}
\end{figure*} 

Figures \ref{fig:alma}, \ref{fig:nir} and \ref{fig:metis} show the simulated observations respectively for ALMA wavelengths (870 $\mu m$ in band 7), in the NIR and in the MIR. 
It is clear how the presence of a gap is readily detectable in some of these images, despite the relatively low mass of the planet. Not only the gap itself, but also the bright ring outside the gap edge contributes in making the gap stand out visually. The comparison between the different ALMA resolutions allows us to establish what is the spatial resolution that allows us to reach the ``intrinsic'' boundary posed by the fact that a minimum planet mass is required to significantly perturb the surface density of the disc. The $12 M_\oplus$ planet is observable at the best resolution of 0.027\arcsec (corresponding to slightly more than 2 scale-heights), but not when the resolution is degraded to 0.091\arcsec. 
We also note that sub-millimetre observations have a slight advantage over that of scattered light imaging when it comes to detecting the signature
of a low-mass planet.
The $12 M_\oplus$ planet, while still visible with the highest angular resolution available to ALMA, is impossible to detect in the scattered light images. We interpret this phenomenon as due to the fact that the sub-mm traces larger particles which are less coupled to the gas surface density. As noted in section \ref{sec:dustsigma}, this means that the gap is more pronounced for the large particles. This results is consistent with what other authors have found \citep{Dong2015}. For what concerns the MIR, we note that the results are largely similar to the NIR scattered light. While the MIR traces slightly larger particles, the difference is not significant enough to affect the images. Therefore the minimum mass of a planet that can be detected
through a gap in the dust image is wavelength dependent. In the sub-mm, we can almost go down to $10 M_\oplus$, while in the NIR and MIR (when it will become possible in the future) we are constrained to slightly larger masses of $\sim 20 M_\oplus$.

\subsection{Varying the aspect ratio}
\begin{figure*}
\includegraphics[width=\textwidth]{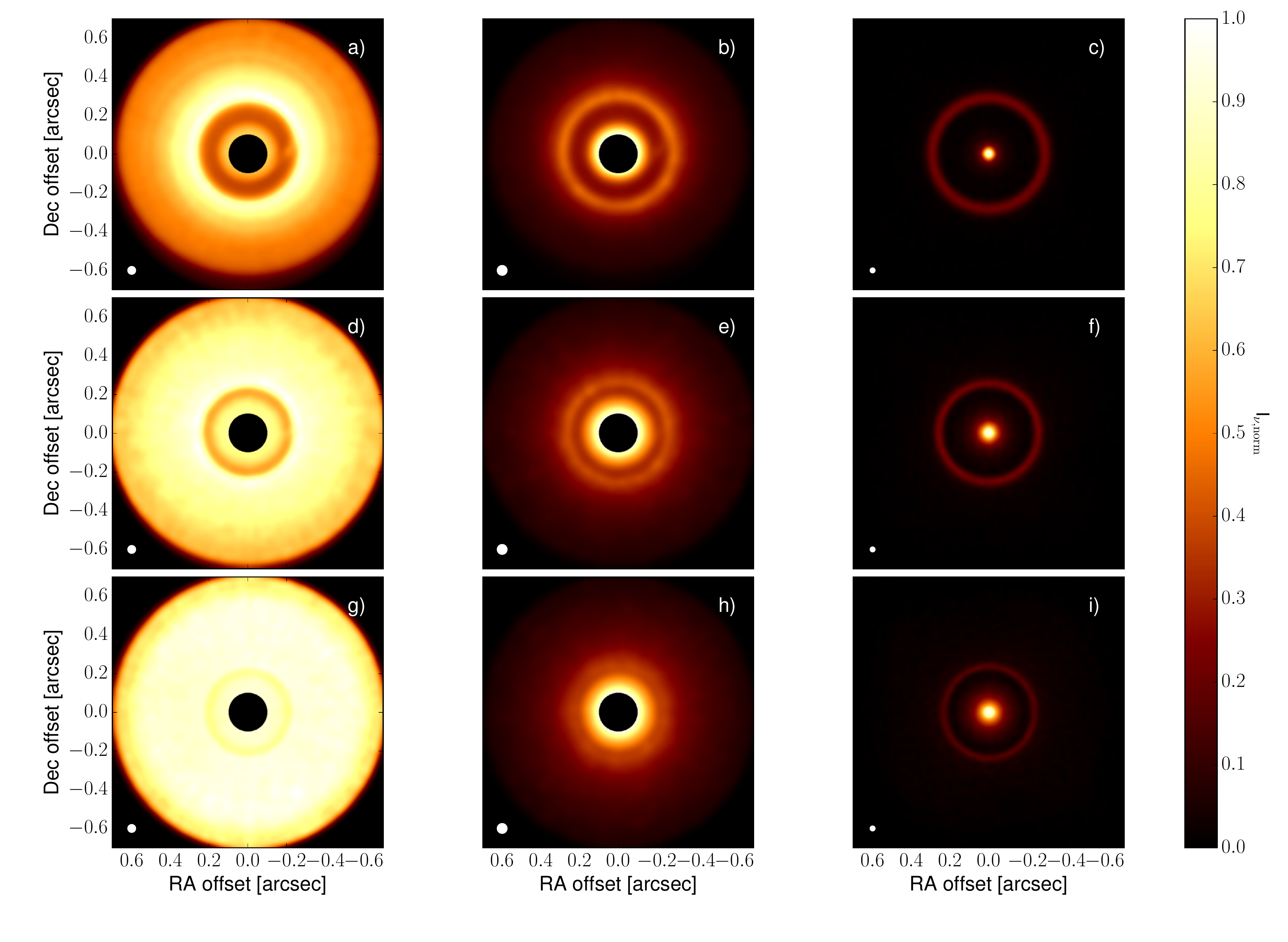}
\caption{An example of how the aspect ratio of the disc changes the morphology of the image. We show the results for a planet mass of $20 M_\oplus$, which corresponds to $2.5 M_\oplus$ for the ``cold'' disc and to $160 M_\oplus$ for the ``hot'' disc once the planet mass has been rescaled to have the same $M_{pl}/M_{gap}$ ratio. The top row is the ``hot'' disc, the middle row is the fiducial model, and the bottom row is the ``cold'' disc. The columns show different instruments: SPHERE (left), METIS (middle) and ALMA (right). In the left and middle column the images are scaled with the square of the radial distance from the star and central 0.1\arcsec of the disc was masked to simulate the effect of a coronograph. The size of the PSF/synthesised beam is shown in the bottom left corner as a filled white ellipse.}
\label{fig:all_aspect}
\end{figure*}

When we change the disc aspect ratio the gap opening mass changes (equation \ref{eq:thermal}). We might therefore expect that a simulation at a given $H/R$ and planet mass should be ``equivalent'' in terms of observability to one with another $H/R$ value and a planet mass scaled according to $(H/R)^3$. This is however not entirely true because the overall spatial scale of the induced structure scales with $H$. Given finite instrumental resolution this means that in cold discs the minimum detectable planet mass does not decrease as steeply with $H/R$ as the cubic dependence suggested by the simple argument above. This is illustrated in figure \ref{fig:all_aspect} where we show simulated observations for the three $H/R$ values we simulate (0.025, 0.05 and 0.1) and planet masses which represent the same ratio of $M_{pl}/M_{gap}$ (that is, 20 $M_\oplus$ for $H/R=0.05$, 2.5 $M_\oplus$ for the ``cold'' disc and 160 $M_\oplus$ for the ``hot'' disc). Although the simulations produce comparable depths of gaps in the gas, the very narrow feature produced by the lowest mass planet in the coldest disc is clearly the most challenging to detect observationally.

Thus the detectability of the gap opened by a planet depends on two factors, the amount of depletion within the gap and the width of the gap. 
If the resolution of the observations is much higher (i.e. the size of the PSF/synthesised beam is much smaller than the width of the gap) the 
detectability of the gap depends only on the S/N of the image (i.e. whether or not the given amount of depletion in the gap could be robustly 
measured above the noise level). If, on the other hand, the width of the gap is comparable or smaller than the size of the PSF / synthesised beam, the contrast between the gap and the surrounding disc is reduced, making the detectability of the gap more difficult. Indeed this is the case in the ``cold'' disc for the 2.5 $M_\oplus$ planet, which creates a gap that is narrower than the PSF.

\section{Discussion}

\label{sec:discussion}

\subsection{Can we measure the planet mass?}

\label{sec:measure_mass}

Supposing that a real disc with the observed morphology described in this paper is found, a very important question is whether it is possible to measure the mass of the putative planet responsible for gap opening. { As mentioned in section \ref{sec:ic}, all the planet masses we quote in this section assume a central star of $1 \ M_\odot$, and need to be rescaled properly with the stellar mass}. We have shown that in this low mass planet regime we do not find any evidence for non axisymmetric structures. This is in contrast to more massive planets that create vortices if the disc has a sufficiently low viscosity \citep{Regaly2012}, an explanation that has been proposed for the asymmetries observed in transition discs.



In terms of \textit{qualitative} differences, thus, the only one present in the cases discussed in this paper is the difference between gaps and holes in the sub-mm dust. The presence of a hole in the sub-mm dust points to a planet mass $\gtrsim 20 M_\oplus$ for the fiducial model ($H/R=0.05$). We can then conclude that the existence of a hole in the sub-mm points to a planet mass $\gtrsim 0.2 M_{th} \sim 20 ((H/R)/0.05)^3 M_\oplus$, which corresponds to the minimum mass for which there is a pressure maximum in the gas. The opposite should however be interpreted with caution. As discussed briefly in section \ref{sec:gassigma}, reaching a steady state takes a considerable time (thousands of orbits). Before the steady state is reached, the planet may be able to open only a gap, which will slowly turn into a hole. We remark that 1000 orbits at 30 AU is approximately $1.6 \times 10^5 \ \mathrm{yr}$, which is a sizeable fraction of the disc lifetime (2-3 $\mathrm{Myr}$, \citealt{Fedele2010}). Therefore, observing a gap does \textit{not} necessarily mean that the planet is less massive than the previously stated threshold.

To get better estimates on the planet mass, it is thus necessary to use \textit{quantitative} arguments. In particular, the diagnostics that have been proposed in the literature \citep{ovelar2013,Kanagawa2015,Akiyama2015} are the gap depth, the gap width and the location of the bright ring in the sub-mm image. 

To quantify the shape and position of the gap, we compute these quantities from our simulated images in the following way. First we de-projected the image with the known inclination and position angles and calculated the azimuthally averaged radial surface brightness profile of the disc.
We note that in reality the inclination of the disc can be constrained fairly well from sub-millimetre line observations. Then we fitted a first order 
polynomial in log-log space to the radial intervals [0.075\arcsec, 0.1\arcsec] and [0.45\arcsec, 0.55\,arcsec] to get a model of the background surface
brightness profile. We then normalised the azimuthally averaged radial surface brightness profile of the disc to the fitted polynomial. Finally
we measured the parameters of the gap on this normalised radial surface brightness profile, I$_{\nu}$, between 0.1\arcsec and 0.45\arcsec.

The depth of the gap was taken to be the minimum of the normalised surface brightness, I$_{\nu, {\rm min}}$. The width of the gap was taken to be the distance between the two radii (inner and outer radius of the gap), where the normalised radial surface brightness dropped below 
(1-$0.66\times$(1-I$_{\nu, {\rm min}}$)). 
The location of the bright ring in the sub-mm was determined as the position of the maximum in the normalised surface brightness 
outside of the gap. It is important to note that we assumed that gaps are detectable if the surface brightness reduction in the gap is more than
or equal to 50\% compared to the fitted background surface brightness of the disc.

Regarding the gap depth as derived from the sub-mm images, we remark that there is only a narrow range of planet masses that is able to create gaps. The other planets either do not affect the dust surface density, or create actual holes, for which measuring the contrast is not meaningful. For this reason we caution against the use of the relations for the gap depth that have been derived for the gas \citep{Kanagawa2015}, as they are unlikely to hold when applied to sub-mm observations and they lead to serious overestimates of the planet mass. Moreover, the gap depth is affected by the finite resolution of the observations, which makes the gap shallower than what it would be in an image with infinite resolution.

\begin{figure}
\includegraphics[width=\columnwidth]{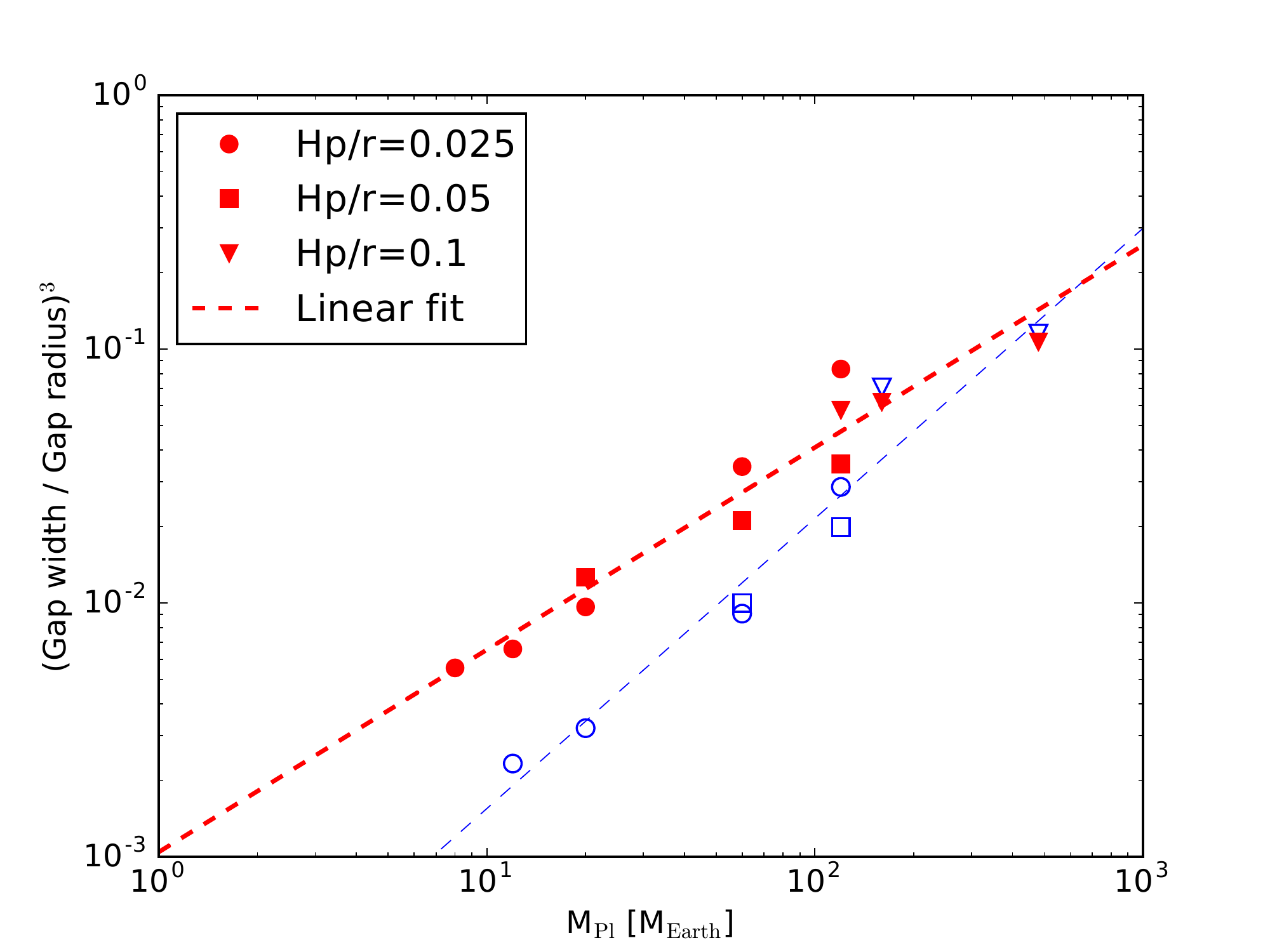}
\caption{Gap width derived from the scattered light observations (normalised to the gap distance from the star, { and to the third power to stress more the dependence on the planet mass}), as a function of the planet mass 
at the time of 400 (blue symbols) and 3000 (red symbols) planetary orbits. The different markers indicate the different aspect ratios and we show a linear fit. While a clear and tight correlation holds between the gap width and the planet mass at any given time the correlation evolves in time, which leads to a factor of a few uncertainty in planet mass if the system is very young. Note that gap width does not depend on the disc temperature, but the gap depth does. {The reason why the points for different aspect ratios cover different ranges in planet masses is that in a warmer disc a higher planet mass is required to open a gap. The planet masses assume a star of 1 $M_\odot$. For stars of different mass, the mass should be rescaled to the same mass ratio accordingly.} }
\label{fig:gap_sigma_NIR}
\end{figure}

The gap width as measured from the sub-mm images suffers from the same problem, namely that for the most massive planets the gaps will slowly turn into holes for which a width is no longer defined. We thus consider in what follows the gap width as measured from the SPHERE NIR images. The results are plotted in Figure \ref{fig:gap_sigma_NIR}. Note, that in order to make the correlation between the planet mass and the gap width 
independent of both the distance to the system and of the distance between the planet and the star we normalised the gap width with the distance of 
the gap centre from the star. { Finally, to stress more the dependence on the planet mass, we show in the figure the cube of the gap width. From the theoretical point of view, this is justified by the fact that we expect the gap width to be set by the Hill radius of the planet, and we thus expect the quantity we plot to scale linearly with the planet mass}. We can see how the gap width is a good estimator of the planet mass. As discussed in section \ref{sec:gassigma}, there is still some time evolution on the timescale of $\sim 10^3$ orbits, and therefore we show the correlation at two different times (400 and 3000 orbits) to bracket the importance of time evolution. A linear fit to the data points in the form of $\log{y} = C_0 + C_1\log{x}$ ({ where the logarithms are in base 10}) results in coefficients of $C_0=-2.981$, $C_1=0.797$ and $C_0=-3.953$, $C_1=1.143$ for 400 and 3000 planetary orbits, respectively. The masses derived from the relations presented here should then be considered only as lower limits in case the gap is created by a more massive planet at earlier times. The uncertainty in the planet mass determination is a factor 2-3, although we stress that, unless the planet is very young, observations are more likely to target planets where the gap width has reached convergence. Note also that there is no need to have any knowledge of the disc temperature to do this plot; quantities that can be derived directly from the observations are sufficient. This is because, { as already mentioned,} the gap width is mostly set by the Hill radius of the planet (see \citealt{2013ApJ...769...41D} for a discussion about the regimes where this holds).

One needs to keep in mind though that, while the gap width does not depend on the disc temperature, the gap depth does, and the same planet in a colder disc creates a deeper gap. Along the same line, in a warmer disc a more massive planet is necessary to create a gap that is deep enough to be observed (which is the reason why the points for different aspect ratios cover different ranges in planet masses).


\begin{figure}
\includegraphics[width=\columnwidth]{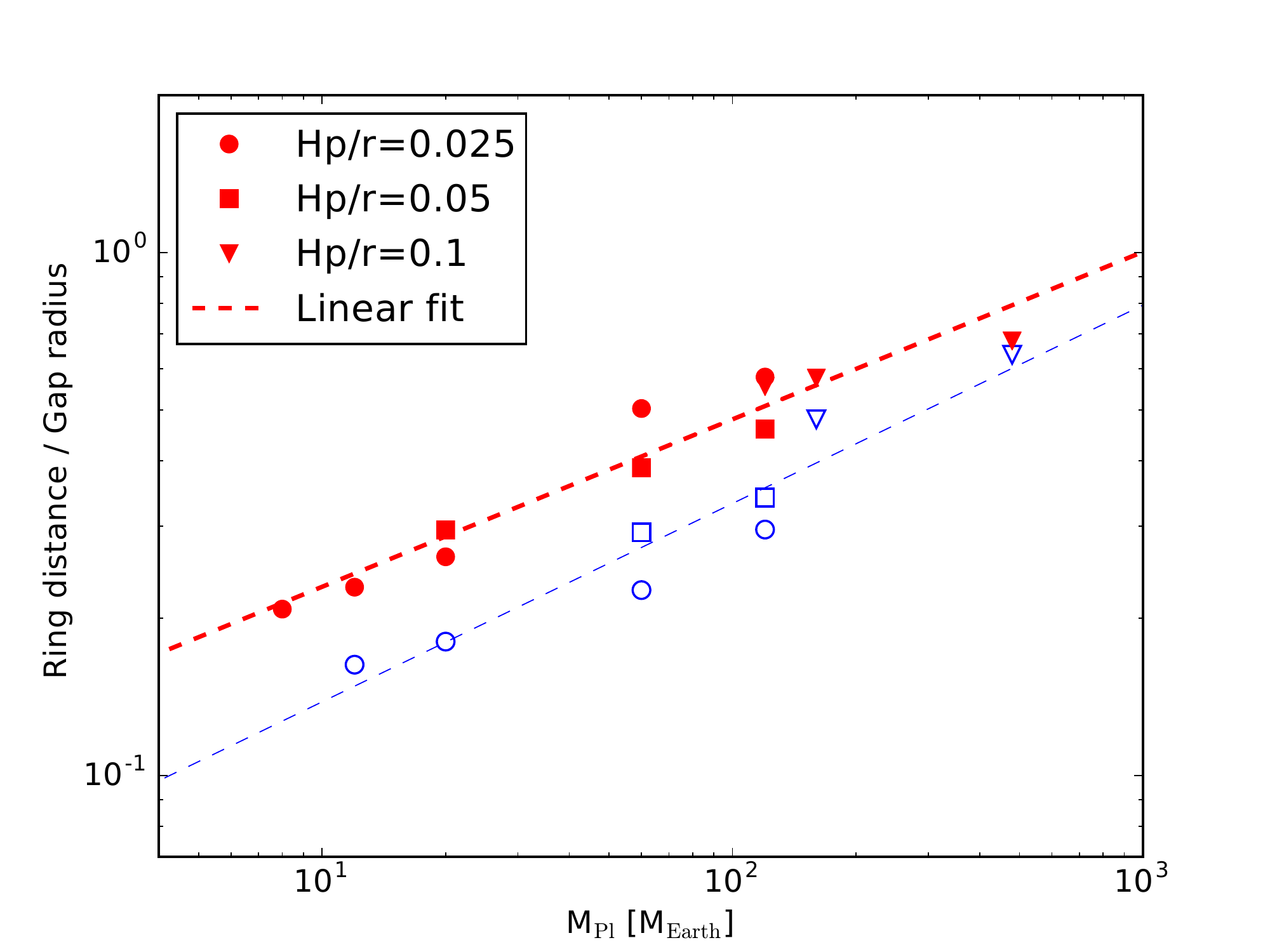}
\caption{Ratio of the radial distance of the pressure maximum outside the planet orbit (as derived from the sub-mm image) from the location of the planet (reconstructed using the center of the gap in the NIR image) to the radius of the gap center as a function of planet mass. The correlation is evident and it holds even when varying the disc aspect ratio (indicated by the marker). However it varies with time as the two sets of points show (blue data points: 400 orbits; red data points: 3000 orbits), so that the planet mass derived from the final correlation can be considered only as a lower limit on the planet mass, as one cannot exclude that the system is very young. { The planet masses assume a star of 1 $M_\odot$. For stars of different mass, the mass should be rescaled to the same mass ratio accordingly.}
}
\label{fig:delta_obs}
\end{figure}

Finally, the third diagnostic that we consider is the radial distance of the maximum in the sub-mm images from the planet location. This is similar to what has been proposed in the context of transition discs \citep{ovelar2013}, namely that the different location of the holes at different wavelengths (normally scattered light in comparison with sub-mm) is the smoking gun for the presence of a pressure maximum, and in theory this effect can be used to measure the planet mass. For the planet masses that we consider we have shown how the scattered light images always show gaps rather than holes. We can therefore use the image to pinpoint the radial location of the planet, which we assume to be at the center of the gap. In transition discs instead one has to resort to use the inner edge of the NIR image (as there might be a hole). We can then use the sub-mm image to find the maximum in the dust surface density, which traces the pressure maximum in the gas. This method thus requires two observations at different wavelengths. While we use NIR and sub-mm in what follows, the requirement for the first wavelength is to show a gap rather than a hole (so that one can constrain the radial position of the planet) and for the second one to exhibit a maximum in the surface brightness at the pressure maximum location. 
In Figure \ref{fig:delta_obs} we show the ratio of the radial distance of the maximum from the location of the planet as a function of the planet mass
estimated from the simulated observations at 400 orbits and at 3000 orbits of the planet.  To make the plot independent of both the distance to the system and the distance of the planet from the star we normalise the radial distance of the maximum from the planet by the distance of the gap centre from the star. As can be seen the correlation between the position of the maximum and the planet mass is clear and tight at each epoch, even though
a clear time evolution is also present in the figure as in the previous case, since the position of the maximum moves further away from the gap with time (see section \ref{sec:gassigma}). As shown in figure \ref{fig:press_max_t}, the position of the maximum has reached converge after 3000 orbits, so that the red points are representative of the subsequent evolution. A linear
fit to the data points in the form of $\log{y} = C_0 + C_1\log{x}$ (({ where, like before, the logarithms are in base 10})) results in coefficients of $C_0=-1.238$, $C_1=0.379$ and $C_0=-0.959$, $C_1=0.320$ for the points at 400 and 3000 planetary orbits, respectively.

Given that we expect the quantities plotted to scale with the intensity of the pressure forces, the existence of a correlation shows that the distance of the pressure maximum must scale with the Hill radius (as is the case for the gap width). Indeed this is confirmed by performing a fit to the distance normalised by the disc scale-height as a function of the planet mass normalised to the gap opening mass, which shows how the distance of the sub-mm ring $\Delta_{\rm mm}$ is $\Delta_{\rm mm} \simeq 10 (M_{\rm pl}/M_{\rm th})^{1/3} \, H = 10 R_H$ after 3000 orbits ({ the same result can be obtained from the coefficients of the fit given above}). Using non-normalised quantities has the advantage, though, that it is not required to know the disc temperature. The same caveats as before about the fact that it is the disc temperature that sets the minimum threshold mass still hold. We also experimented with, instead of using the simulated observations, deriving the same quantity from the gas surface density, where we can also measure the locations of the planet and of the pressure maximum. No noticeable difference is found { for the pressure maximum position}, which shows that this quantity is a very robust indicator of the planet mass as from \textit{dust} (continuum) observations one can reconstruct fundamental properties of the (unseen) \textit{gas} disc. { Our method instead systematically slightly underestimates the position of the planet, which we attribute to radiative transfer effects. The error doubles in the final estimate as we normalise the distance to the gap center. Using directly the values from the simulations yields then a slightly different coefficient $C_0 \simeq -1.1$, or  $\Delta_{\rm mm} \simeq 7.5 R_H$.}

Of the two criteria that we discussed, the gap width has the advantage of requiring observations only at one wavelength. However, it needs to be borne in mind that scattered light observations have a slightly higher threshold mass for observability. Moreover, experimenting with deriving a gap width has shown us how the gap width is somehow sensitive to the exact way the fit is done, and to the details of the gap shape. The position of the pressure maximum has instead the advantage of being straightforward to apply and requiring little manipulation of the data.




\subsection{Prospects for the future}
While currently there is no instruments in operation in the MIR with sufficiently high resolution for detecting low mass planets, the 30-m telescope class will be a great improvement. It will provide the same resolution we have currently with SPHERE or ALMA. This work shows that the thermal emission of the disc in the MIR is a good wavelength to use to probe the structures sculpted by planets in the disc.

However the hard boundary in the planet mass that we discuss in this paper is set by an intrinsic property, that is, how strong is the perturbation in the dust, rather than by a limitation in angular resolution. The only requirement on the angular resolution is that is sufficient to detect the presence of a gap (if it is there). While the resolutions we considered in this paper are enough for reaching the ``intrinsic'' boundary (i.e. resolving the gap width for a planet of the minimum mass required to open a gap in the dust) at 30 AU, improvements in spatial resolution in the future will still allow us to observe planets closer to the star { or in discs further from Earth}.

\subsection{Limitations and other effects}

\label{sec:limitation}

We assumed the planet is at several tens of AU. Although very little is known about planets at these locations (with only a handful of planets around main sequence stars detected by direct imaging, \citealt{Chauvin2014}), this choice is backed up by an image like HL Tau. Although other interpretations for the gap structure (which do not invoke the presence of planets) have been proposed so far \citep{2015ApJ...806L...7Z}, if planets are responsible for the observed morphology, it means that they can form at these large orbital distances and be found in proto-planetary discs. Recent observations of LkCa15 \citep{2015Natur.527..342S} also point in the same direction; moreover, a common explanation of transition discs (see \citealt{Espaillat2014} for a review) is the presence of massive planets at tens of AU. The presence of a super-Earth at tens of AU is particularly challenging to explain by theoretical models as in core accretion the timescale for forming planets at these orbital distances is longer than the disc lifetime (but see \citealt{2012A&A...544A..32L} for an alternative), while gravitational instability tends to form much more massive objects. A possibility \citep{2013ApJ...775...42I,2014ApJ...797....1K} is that they have been scattered by giant planets that formed closer to the star. If they manage to accrete gas, they might eventually turn into very massive giants ($\gtrsim 10 M_j$) given that most of the mass is in the outer part of the disc.

In this work we assumed a constant Stokes number - we did not include dust coagulation and fragmentation. Due to numerical limitations, no work so far has implemented a full dust evolution model on top of a hydro simulation, although approximate attempts have been done \citep{2015MNRAS.454L..36G}, or just static snapshots from the hydro simulation have been used \citep{Pinilla2012,ovelar2013}. This is an aspect that is worth exploring in future works given that the timescales that we have run our simulations for are a sizeable fraction of the disc lifetime. In addition it has been proved that dust evolution alone can create gaps in scattered light images \citep{2015ApJ...813L..14B}, and therefore is important to disentangle the two effects. Our implementation also contains no feedback of the dust onto the gas. However, this is a minor limitation for the lowest mass planets we simulate (which are the real focus of this paper) given the modest amount of dust pile up at the outer edge of the gap.

Our simulations are not 3D, a limitation that we share with other works that have been done in this area \citep{Zhu2012,Dong2015,Picogna2015}. This is unlikely to affect the big grains, whose dynamics is limited to the midplane due to settling. In addition, at long wavelengths discs are optically thin so that observations probe the disc midplane. If there is a vertical dependency of the dynamics, it might however affect the small dust since observations at short wavelengths (i.e. in scattered light) probe only the surface of the disc. In particular, the temperature and the stopping time depend on the height above the midplane - in addition there might be vertical motions \citep[e.g.,][]{2014Icar..232..266M,2015ApJ...811..101F} that complicate the picture even more. This should be explored in future papers. We remark in particular that, of the two ways of estimating the planet mass that we discussed, this affects mostly the NIR gap width, which therefore we regard as more uncertain. Given that the other method we propose instead uses the NIR image only to pinpoint at the radial location of the planet, we expect it to be resilient to potential 3D effects.

We did not include migration. Taken at face value, the relevant migration rate, type I, is very fast \citep{2002ApJ...565.1257T}, so that our approximation might not appear fully justified. However it is known that such rate is not compatible with the observed properties of exo-planetary systems \citep{2008ApJ...673..487I}, and a reduction in the migration rate of almost two orders of magnitude is needed to reconcile with exoplanet statistics. While there is considerable debate on the physical reasons \citep[e.g.,][]{2006ApJ...642..478M,2011A&A...536A..77B,2011MNRAS.417.1236H}, all explanations require additional physics which is not present in our simulation. Therefore, in light of these results it is safe to say that the migration rate as computed in our locally isothermal simulation would not be the correct one anyway. We plan to explore this effect too in future papers. { Finally, it should be added that recent research \citep[e.g.,][]{2010ApJ...715L..68L,2014A&A...570A..75B} has also highlighted the existence of zero-torque radii where the type I migration rate vanish. If the planet is at one of those locations then neglecting migration as we have done in this paper is justified.}


Finally, we were limited to study dust smaller than $St=1$ by the choice of modelling dust as a fluid. In practice, given that the planets that we simulate in this paper only open shallow gaps, this is not a real limitation. Using particles rather than a fluid we could study bigger dust, but dust diffusion, which we showed to be an important contribution in setting the dust radial velocity, would have been considerably more difficult to implement.

\section{Conclusions}

\label{sec:concl}

In this paper we have undertaken a systematic study to establish how the observational signatures of low mass planets embedded in proto-planetary discs depend on the disc properties. We can conclude that:

\begin{itemize}
\item Low mass planets (tens of $M_\oplus$), even if they are not fully in the gap opening regime, can generate observational signatures in protoplanetary discs, consisting in a gap at the radial location of the planet and a bright ring at the gap edge.
\item The observational signatures are always azimuthally symmetric.
\item In terms of current facilities, we find that the observational signatures can be traced both in the sub-mm (with ALMA) and in the scattered light (with instruments such as SPHERE or GPI). We have highlighted how the observational signatures are present also in the MIR, which will become observable with the upcoming generation of 30m class telescopes.
\item For an aspect ratio $H/R$ of 0.05, the threshold to perturb the dust surface density is $\sim 15 M_\oplus$, with some dependence on the wavelength of the observations which favours the sub-mm. In particular, the threshold is closer to $\sim 12 M_\oplus$ in the sub-mm and $\sim 20 M_\oplus$ in the NIR or MIR. More importantly, though, the threshold is highly dependent (roughly with the third power) on the disc aspect ratio. We predict thus that the minimum mass is $ \sim 15 ((H/R)/0.05)^3 M_\oplus$. This is an \textit{intrinsic} boundary.
\item Real observations are limited though by the finite resolution. We show that at the fiducial distance of 30 au from the star, the threshold for detection with existing instruments in the NIR and sub-mm and planned instruments in the MIR is set by the intrinsic properties of the disc. Improvements in the telescope capabilities will allow to sample regions closer to the star.
\item We { confirm} \citep{2014A&A...572A..35L} that a planet mass $\gtrsim 0.2 M_{th} \sim 20 ((H/R)/0.05)^3 M_\oplus$ is required to create a pressure maximum in the gas at the gap outer edge. After enough time has passed, the inescapable consequence for these planets is to create a hole in the sub-mm dust. Planets smaller than this threshold can still affect the dust significantly and create a gap in the sub-mm, although the relevant range of planet masses is rather limited (a planet 2.5 times less massive than the quoted threshold does not produce any observational signature).
\item It is interesting also to explore the inverse problem, that is, to derive the planet mass from high resolution observations. Using the gap width or contrast in the sub-mm (ALMA) images to measure the planet mass is complicated by the fact that the gaps will eventually turn into holes over a wide range of planet masses. We thus disfavour this method to measure the planet mass and prefer to use other diagnostics.
\item We find that the gap width as derived from the scattered light images is a good indicator of the planet mass. This indicator is affected by time evolution and it takes roughly 1000 orbits (which at 30 AU corresponds to $\sim 1.6 \times 10^5 \, \mathrm{yr}$) to reach the final value. If the system is very young, this might lead to a factor 2-3 underestimate in the planet mass determination. Additionally, the exact definition of gap width is sensitive to the details of the gap shape.
\item We find also the location of the bright ring in the sub-mm images, which traces the gas pressure maximum, to be a robust indicator of the planet mass. In case the sub-mm image shows a hole it is necessary to use the scattered light image to pinpoint the radial location of the planet. We expect this method to be the most resilient to the details of the dust dynamics; the same caveats about time dependence as before apply.

\end{itemize}

\section*{Acknowledgements}
We thank an anonymous referee for a careful reading of our manuscript and many useful comments. We thank Leonardo Testi for a stimulating discussion that started this work, Sijme-Jan Paardekooper and Richard Alexander for their constructive criticism, Judith Ngoumou and the Munich Star Formation Coffee for a very lively discussion. This work has been supported by the DISCSIM project, grant
agreement 341137 funded by the European Research Council under ERC-2013-ADG. This work used the DIRAC Shared Memory Processing system at the University of Cambridge, operated by the COSMOS Project at the Department of Applied Mathematics and Theoretical Physics on behalf of the STFC DiRAC HPC Facility (www.dirac.ac.uk). This equipment was funded by BIS National E-infrastructure capital grant ST/J005673/1, STFC capital grant ST/H008586/1, and STFC DiRAC Operations grant ST/K00333X/1. DiRAC is part of the National E-Infrastructure.

\bibliography{BiblioRosotti}{}

\begin{thebibliography}{}

\bibitem[\protect\citeauthoryear{{Akiyama}, {Hasegawa}, {Hayashi} \&
  {Iguchi}}{{Akiyama} et~al.}{2015}]{Akiyama2015}
{Akiyama} E.,  {Hasegawa} Y.,  {Hayashi} M.,    {Iguchi} S.,  2015, ArXiv
  e-prints

\bibitem[\protect\citeauthoryear{{ALMA Partnership}, {Brogan}, {P{\'e}rez},
  {Hunter}, {Dent}, {Hales}, {Hills}, {Corder}, {Fomalont}, {Vlahakis} \&
  {Asaki}}{{ALMA Partnership} et~al.}{2015}]{hltau}
{ALMA Partnership} {Brogan} C.~L.,  {P{\'e}rez} L.~M.,  {Hunter} T.~R.,  {Dent}
  W.~R.~F.,  {Hales} A.~S.,  {Hills} R.~E.,  {Corder} S.,  {Fomalont} E.~B.,
  {Vlahakis} C.,    {Asaki} Y.,  2015, \apjl, 808, L3

\bibitem[\protect\citeauthoryear{{Baruteau}, {Crida}, {Paardekooper}, {Masset},
  {Guilet}, {Bitsch}, {Nelson}, {Kley} \& {Papaloizou}}{{Baruteau}
  et~al.}{2014}]{2014prpl.conf..667B}
{Baruteau} C.,  {Crida} A.,  {Paardekooper} S.-J.,  {Masset} F.,  {Guilet} J.,
  {Bitsch} B.,  {Nelson} R.,  {Kley} W.,    {Papaloizou} J.,  2014, Protostars
  and Planets VI, pp 667--689

\bibitem[\protect\citeauthoryear{{Benisty}, {Juhasz}, {Boccaletti}, {Avenhaus},
  {Milli}, {Thalmann}, {Dominik}, {Pinilla}, {Buenzli}, {Pohl} \&
  {Beuzit}}{{Benisty} et~al.}{2015}]{2015A&A...578L...6B}
{Benisty} M.,  {Juhasz} A.,  {Boccaletti} A.,  {Avenhaus} H.,  {Milli} J.,
  {Thalmann} C.,  {Dominik} C.,  {Pinilla} P.,  {Buenzli} E.,  {Pohl} A.,
  {Beuzit} J.-L.,  2015, \aap, 578, L6

\bibitem[\protect\citeauthoryear{{Ben{\'{\i}}tez-Llambay}, {Masset},
  {Koenigsberger} \& {Szul{\'a}gyi}}{{Ben{\'{\i}}tez-Llambay}
  et~al.}{2015}]{2015Natur.520...63B}
{Ben{\'{\i}}tez-Llambay} P.,  {Masset} F.,  {Koenigsberger} G.,
  {Szul{\'a}gyi} J.,  2015, \nat, 520, 63

\bibitem[\protect\citeauthoryear{{Beuzit}, {Feldt}, {Dohlen}, {Mouillet},
  {Puget}, {Wildi}, {Abe}, {Antichi}, {Baruffolo}, {Baudoz}, {Boccaletti},
  {Carbillet}, {Charton}, {Claudi}, {Downing}, {Fabron}, {Feautrier} \&
  {Fedrigo}}{{Beuzit} et~al.}{2008}]{2008SPIE.7014E..18B}
{Beuzit} J.-L.,  {Feldt} M.,  {Dohlen} K.,  {Mouillet} D.,  {Puget} P.,
  {Wildi} F.,  {Abe} L.,  {Antichi} J.,  {Baruffolo} A.,  {Baudoz} P.,
  {Boccaletti} A.,  {Carbillet} M.,  {Charton} J.,  {Claudi} R.,  {Downing} M.,
   {Fabron} C.,  {Feautrier} P.,    {Fedrigo} E.,  2008, in Ground-based and
  Airborne Instrumentation for Astronomy II Vol.~7014 of \procspie, {SPHERE: a
  'Planet Finder' instrument for the VLT}.
p. 701418

\bibitem[\protect\citeauthoryear{{Birnstiel}, {Andrews}, {Pinilla} \&
  {Kama}}{{Birnstiel} et~al.}{2015}]{2015ApJ...813L..14B}
{Birnstiel} T.,  {Andrews} S.~M.,  {Pinilla} P.,    {Kama} M.,  2015, \apjl,
  813, L14

\bibitem[\protect\citeauthoryear{{Bitsch} \& {Kley}}{{Bitsch} \&
  {Kley}}{2011}]{2011A&A...536A..77B}
{Bitsch} B.,  {Kley} W.,  2011, \aap, 536, A77

\bibitem[\protect\citeauthoryear{{Bitsch}, {Morbidelli}, {Lega}, {Kretke} \&
  {Crida}}{{Bitsch} et~al.}{2014}]{2014A&A...570A..75B}
{Bitsch} B.,  {Morbidelli} A.,  {Lega} E.,  {Kretke} K.,    {Crida} A.,  2014,
  \aap, 570, A75

\bibitem[\protect\citeauthoryear{{Chauvin}, {Vigan}, {Bonnefoy}, {Desidera},
  {Bonavita}, {Mesa}, {Boccaletti}, {Buenzli}, {Carson} \& {Delorme}}{{Chauvin}
  et~al.}{2014}]{Chauvin2014}
{Chauvin} G.,  {Vigan} A.,  {Bonnefoy} M.,  {Desidera} S.,  {Bonavita} M.,
  {Mesa} D.,  {Boccaletti} A.,  {Buenzli} E.,  {Carson} J.,    {Delorme} P.,
  2014, ArXiv e-prints

\bibitem[\protect\citeauthoryear{{Clarke} \& {Pringle}}{{Clarke} \&
  {Pringle}}{1988}]{ClarkeDiffusion}
{Clarke} C.~J.,  {Pringle} J.~E.,  1988, \mnras, 235, 365

\bibitem[\protect\citeauthoryear{{Crida}, {Morbidelli} \& {Masset}}{{Crida}
  et~al.}{2006}]{CridaOpeningCriterion}
{Crida} A.,  {Morbidelli} A.,    {Masset} F.,  2006, \icarus, 181, 587

\bibitem[\protect\citeauthoryear{{de Juan Ovelar}, {Min}, {Dominik},
  {Thalmann}, {Pinilla}, {Benisty} \& {Birnstiel}}{{de Juan Ovelar}
  et~al.}{2013}]{ovelar2013}
{de Juan Ovelar} M.,  {Min} M.,  {Dominik} C.,  {Thalmann} C.,  {Pinilla} P.,
  {Benisty} M.,    {Birnstiel} T.,  2013, \aap, 560, A111

\bibitem[\protect\citeauthoryear{{de Val-Borro}, {Edgar}, {Artymowicz},
  {Ciecielag}, {Cresswell}, {D'Angelo}, {Delgado-Donate}, {Dirksen}, {Fromang},
  {Gawryszczak} \& {Klahr}}{{de Val-Borro} et~al.}{2006}]{ValBorro2006}
{de Val-Borro} M.,  {Edgar} R.~G.,  {Artymowicz} P.,  {Ciecielag} P.,
  {Cresswell} P.,  {D'Angelo} G.,  {Delgado-Donate} E.~J.,  {Dirksen} G.,
  {Fromang} S.,  {Gawryszczak} A.,    {Klahr} H.,  2006, \mnras, 370, 529

\bibitem[\protect\citeauthoryear{{Dipierro}, {Laibe}, {Price} \&
  {Lodato}}{{Dipierro} et~al.}{2016}]{2016arXiv160207457D}
{Dipierro} G.,  {Laibe} G.,  {Price} D.~J.,    {Lodato} G.,  2016, ArXiv
  e-prints

\bibitem[\protect\citeauthoryear{{Dipierro}, {Price}, {Laibe}, {Hirsh},
  {Cerioli} \& {Lodato}}{{Dipierro} et~al.}{2015}]{Dipierro2015}
{Dipierro} G.,  {Price} D.,  {Laibe} G.,  {Hirsh} K.,  {Cerioli} A.,
  {Lodato} G.,  2015, \mnras, 453, L73

\bibitem[\protect\citeauthoryear{{Dong}, {Zhu} \& {Whitney}}{{Dong}
  et~al.}{2015}]{Dong2015}
{Dong} R.,  {Zhu} Z.,    {Whitney} B.,  2015, \apj, 809, 93

\bibitem[\protect\citeauthoryear{{Duffell} \& {MacFadyen}}{{Duffell} \&
  {MacFadyen}}{2013}]{2013ApJ...769...41D}
{Duffell} P.~C.,  {MacFadyen} A.~I.,  2013, \apj, 769, 41

\bibitem[\protect\citeauthoryear{{Espaillat}, {Muzerolle}, {Najita}, {Andrews},
  {Zhu}, {Calvet}, {Kraus}, {Hashimoto}, {Kraus} \& {D'Alessio}}{{Espaillat}
  et~al.}{2014}]{Espaillat2014}
{Espaillat} C.,  {Muzerolle} J.,  {Najita} J.,  {Andrews} S.,  {Zhu} Z.,
  {Calvet} N.,  {Kraus} S.,  {Hashimoto} J.,  {Kraus} A.,    {D'Alessio} P.,
  2014, ArXiv e-prints

\bibitem[\protect\citeauthoryear{{Fedele}, {van den Ancker}, {Henning},
  {Jayawardhana} \& {Oliveira}}{{Fedele} et~al.}{2010}]{Fedele2010}
{Fedele} D.,  {van den Ancker} M.~E.,  {Henning} T.,  {Jayawardhana} R.,
  {Oliveira} J.~M.,  2010, \aap, 510, A72

\bibitem[\protect\citeauthoryear{{Fouchet}, {Gonzalez} \& {Maddison}}{{Fouchet}
  et~al.}{2010}]{2010A&A...518A..16F}
{Fouchet} L.,  {Gonzalez} J.-F.,    {Maddison} S.~T.,  2010, \aap, 518, A16

\bibitem[\protect\citeauthoryear{{Fung}, {Artymowicz} \& {Wu}}{{Fung}
  et~al.}{2015}]{2015ApJ...811..101F}
{Fung} J.,  {Artymowicz} P.,    {Wu} Y.,  2015, \apj, 811, 101

\bibitem[\protect\citeauthoryear{{Fung}, {Shi} \& {Chiang}}{{Fung}
  et~al.}{2014}]{2014ApJ...782...88}
{Fung} J.,  {Shi} J.-M.,    {Chiang} E.,  2014, \apj, 782, 88

\bibitem[\protect\citeauthoryear{{Garaud}, {Barri{\`e}re-Fouchet} \&
  {Lin}}{{Garaud} et~al.}{2004}]{Garaud2004}
{Garaud} P.,  {Barri{\`e}re-Fouchet} L.,    {Lin} D.~N.~C.,  2004, \apj, 603,
  292

\bibitem[\protect\citeauthoryear{{Garufi}, {Quanz}, {Avenhaus}, {Buenzli},
  {Dominik}, {Meru}, {Meyer}, {Pinilla}, {Schmid} \& {Wolf}}{{Garufi}
  et~al.}{2013}]{Garufi2013}
{Garufi} A.,  {Quanz} S.~P.,  {Avenhaus} H.,  {Buenzli} E.,  {Dominik} C.,
  {Meru} F.,  {Meyer} M.~R.,  {Pinilla} P.,  {Schmid} H.~M.,    {Wolf} S.,
  2013, \aap, 560, A105

\bibitem[\protect\citeauthoryear{{Gonzalez}, {Laibe}, {Maddison}, {Pinte} \&
  {M{\'e}nard}}{{Gonzalez} et~al.}{2015}]{2015MNRAS.454L..36G}
{Gonzalez} J.-F.,  {Laibe} G.,  {Maddison} S.~T.,  {Pinte} C.,    {M{\'e}nard}
  F.,  2015, \mnras, 454, L36

\bibitem[\protect\citeauthoryear{{Gonzalez}, {Pinte}, {Maddison}, {M{\'e}nard}
  \& {Fouchet}}{{Gonzalez} et~al.}{2012}]{2012A&A...547A..58G}
{Gonzalez} J.-F.,  {Pinte} C.,  {Maddison} S.~T.,  {M{\'e}nard} F.,
  {Fouchet} L.,  2012, \aap, 547, A58

\bibitem[\protect\citeauthoryear{{Hasegawa} \& {Pudritz}}{{Hasegawa} \&
  {Pudritz}}{2011}]{2011MNRAS.417.1236H}
{Hasegawa} Y.,  {Pudritz} R.~E.,  2011, \mnras, 417, 1236

\bibitem[\protect\citeauthoryear{{Howard}, {Marcy}, {Bryson}, {Jenkins},
  {Rowe}, {Batalha}, {Borucki}, {Koch}, {Dunham}, {Gautier} III \& {Van
  Cleve}}{{Howard} et~al.}{2012}]{Howard2012}
{Howard} A.~W.,  {Marcy} G.~W.,  {Bryson} S.~T.,  {Jenkins} J.~M.,  {Rowe}
  J.~F.,  {Batalha} N.~M.,  {Borucki} W.~J.,  {Koch} D.~G.,  {Dunham} E.~W.,
  {Gautier} III T.~N.,    {Van Cleve} J.,  2012, \apjs, 201, 15

\bibitem[\protect\citeauthoryear{{Ida} \& {Lin}}{{Ida} \&
  {Lin}}{2008}]{2008ApJ...673..487I}
{Ida} S.,  {Lin} D.~N.~C.,  2008, \apj, 673, 487

\bibitem[\protect\citeauthoryear{{Ida}, {Lin} \& {Nagasawa}}{{Ida}
  et~al.}{2013}]{2013ApJ...775...42I}
{Ida} S.,  {Lin} D.~N.~C.,    {Nagasawa} M.,  2013, \apj, 775, 42

\bibitem[\protect\citeauthoryear{{Jin}, {Li}, {Isella}, {Li} \& {Ji}}{{Jin}
  et~al.}{2016}]{2016ApJ...818...76J}
{Jin} S.,  {Li} S.,  {Isella} A.,  {Li} H.,    {Ji} J.,  2016, \apj, 818, 76

\bibitem[\protect\citeauthoryear{{Juh{\'a}sz}, {Benisty}, {Pohl}, {Dullemond},
  {Dominik} \& {Paardekooper}}{{Juh{\'a}sz} et~al.}{2015}]{AttilaSpirals}
{Juh{\'a}sz} A.,  {Benisty} M.,  {Pohl} A.,  {Dullemond} C.~P.,  {Dominik} C.,
    {Paardekooper} S.-J.,  2015, \mnras, 451, 1147

\bibitem[\protect\citeauthoryear{{Kanagawa}, {Muto}, {Tanaka}, {Tanigawa},
  {Takeuchi}, {Tsukagoshi} \& {Momose}}{{Kanagawa} et~al.}{2015}]{Kanagawa2015}
{Kanagawa} K.~D.,  {Muto} T.,  {Tanaka} H.,  {Tanigawa} T.,  {Takeuchi} T.,
  {Tsukagoshi} T.,    {Momose} M.,  2015, \apjl, 806, L15

\bibitem[\protect\citeauthoryear{{Kenyon} \& {Hartmann}}{{Kenyon} \&
  {Hartmann}}{1995}]{KenyonHartmann95}
{Kenyon} S.~J.,  {Hartmann} L.,  1995, \apjs, 101, 117

\bibitem[\protect\citeauthoryear{{Kikuchi}, {Higuchi} \& {Ida}}{{Kikuchi}
  et~al.}{2014}]{2014ApJ...797....1K}
{Kikuchi} A.,  {Higuchi} A.,    {Ida} S.,  2014, \apj, 797, 1

\bibitem[\protect\citeauthoryear{{Lambrechts} \& {Johansen}}{{Lambrechts} \&
  {Johansen}}{2012}]{2012A&A...544A..32L}
{Lambrechts} M.,  {Johansen} A.,  2012, \aap, 544, A32

\bibitem[\protect\citeauthoryear{{Lambrechts}, {Johansen} \&
  {Morbidelli}}{{Lambrechts} et~al.}{2014}]{2014A&A...572A..35L}
{Lambrechts} M.,  {Johansen} A.,    {Morbidelli} A.,  2014, \aap, 572, A35

\bibitem[\protect\citeauthoryear{{Lin} \& {Papaloizou}}{{Lin} \&
  {Papaloizou}}{1979}]{GapOpening}
{Lin} D.~N.~C.,  {Papaloizou} J.,  1979, \mnras, 186, 799

\bibitem[\protect\citeauthoryear{{Lin} \& {Papaloizou}}{{Lin} \&
  {Papaloizou}}{1993}]{LinPapaloizou93}
{Lin} D.~N.~C.,  {Papaloizou} J.~C.~B.,  1993, in {Levy} E.~H.,  {Lunine}
  J.~I.,  eds, Protostars and Planets III {On the tidal interaction between
  protostellar disks and companions}.
pp 749--835

\bibitem[\protect\citeauthoryear{{Lyra}, {Paardekooper} \& {Mac Low}}{{Lyra}
  et~al.}{2010}]{2010ApJ...715L..68L}
{Lyra} W.,  {Paardekooper} S.-J.,    {Mac Low} M.-M.,  2010, \apjl, 715, L68

\bibitem[\protect\citeauthoryear{{Masset}}{{Masset}}{2000}]{Masset2000}
{Masset} F.,  2000, \aaps, 141, 165

\bibitem[\protect\citeauthoryear{{Masset}, {Morbidelli}, {Crida} \&
  {Ferreira}}{{Masset} et~al.}{2006}]{2006ApJ...642..478M}
{Masset} F.~S.,  {Morbidelli} A.,  {Crida} A.,    {Ferreira} J.,  2006, \apj,
  642, 478

\bibitem[\protect\citeauthoryear{{Mayor} \& {Queloz}}{{Mayor} \&
  {Queloz}}{1995}]{1995exoplanet}
{Mayor} M.,  {Queloz} D.,  1995, \nat, 378, 355

\bibitem[\protect\citeauthoryear{{Morbidelli}, {Lambrechts}, {Jacobson} \&
  {Bitsch}}{{Morbidelli} et~al.}{2015}]{2015Icar..258..418M}
{Morbidelli} A.,  {Lambrechts} M.,  {Jacobson} S.,    {Bitsch} B.,  2015,
  \icarus, 258, 418

\bibitem[\protect\citeauthoryear{{Morbidelli}, {Szul{\'a}gyi}, {Crida}, {Lega},
  {Bitsch}, {Tanigawa} \& {Kanagawa}}{{Morbidelli}
  et~al.}{2014}]{2014Icar..232..266M}
{Morbidelli} A.,  {Szul{\'a}gyi} J.,  {Crida} A.,  {Lega} E.,  {Bitsch} B.,
  {Tanigawa} T.,    {Kanagawa} K.,  2014, \icarus, 232, 266

\bibitem[\protect\citeauthoryear{{Muto}, {Grady}, {Hashimoto}, {Fukagawa},
  {Hornbeck}, {Sitko}, {Russell}, {Werren}, {Cur{\'e}}, {Currie} \&
  {Ohashi}}{{Muto} et~al.}{2012}]{2012ApJ...748L..22M}
{Muto} T.,  {Grady} C.~A.,  {Hashimoto} J.,  {Fukagawa} M.,  {Hornbeck} J.~B.,
  {Sitko} M.,  {Russell} R.,  {Werren} C.,  {Cur{\'e}} M.,  {Currie} T.,
  {Ohashi} N.,  2012, \apjl, 748, L22

\bibitem[\protect\citeauthoryear{{Owen}}{{Owen}}{2014}]{OwenRadPress}
{Owen} J.~E.,  2014, \apj, 789, 59

\bibitem[\protect\citeauthoryear{{Owen} \& {Wu}}{{Owen} \&
  {Wu}}{2013}]{2013ApJ...775..105O}
{Owen} J.~E.,  {Wu} Y.,  2013, \apj, 775, 105

\bibitem[\protect\citeauthoryear{{Paardekooper} \& {Mellema}}{{Paardekooper} \&
  {Mellema}}{2004}]{paardekooper2004}
{Paardekooper} S.-J.,  {Mellema} G.,  2004, \aap, 425, L9

\bibitem[\protect\citeauthoryear{{Paardekooper} \& {Mellema}}{{Paardekooper} \&
  {Mellema}}{2006}]{PaardekooperMellema2006}
{Paardekooper} S.-J.,  {Mellema} G.,  2006, \aap, 453, 1129

\bibitem[\protect\citeauthoryear{{Picogna} \& {Kley}}{{Picogna} \&
  {Kley}}{2015}]{Picogna2015}
{Picogna} G.,  {Kley} W.,  2015, ArXiv e-prints

\bibitem[\protect\citeauthoryear{{Pinilla}, {Benisty} \& {Birnstiel}}{{Pinilla}
  et~al.}{2012}]{Pinilla12}
{Pinilla} P.,  {Benisty} M.,    {Birnstiel} T.,  2012, \aap, 545, A81

\bibitem[\protect\citeauthoryear{{Pinilla}, {Birnstiel}, {Ricci}, {Dullemond},
  {Uribe}, {Testi} \& {Natta}}{{Pinilla} et~al.}{2012}]{Pinilla2012}
{Pinilla} P.,  {Birnstiel} T.,  {Ricci} L.,  {Dullemond} C.~P.,  {Uribe} A.~L.,
   {Testi} L.,    {Natta} A.,  2012, \aap, 538, A114

\bibitem[\protect\citeauthoryear{{Pollack}, {Hubickyj}, {Bodenheimer},
  {Lissauer}, {Podolak} \& {Greenzweig}}{{Pollack} et~al.}{1996}]{Pollack96}
{Pollack} J.~B.,  {Hubickyj} O.,  {Bodenheimer} P.,  {Lissauer} J.~J.,
  {Podolak} M.,    {Greenzweig} Y.,  1996, \icarus, 124, 62

\bibitem[\protect\citeauthoryear{{Raymond}, {Kokubo}, {Morbidelli}, {Morishima}
  \& {Walsh}}{{Raymond} et~al.}{2014}]{2014prpl.conf..595R}
{Raymond} S.~N.,  {Kokubo} E.,  {Morbidelli} A.,  {Morishima} R.,    {Walsh}
  K.~J.,  2014, Protostars and Planets VI, pp 595--618

\bibitem[\protect\citeauthoryear{{Reg{\'a}ly}, {Juh{\'a}sz}, {S{\'a}ndor} \&
  {Dullemond}}{{Reg{\'a}ly} et~al.}{2012}]{Regaly2012}
{Reg{\'a}ly} Z.,  {Juh{\'a}sz} A.,  {S{\'a}ndor} Z.,    {Dullemond} C.~P.,
  2012, \mnras, 419, 1701

\bibitem[\protect\citeauthoryear{{Rice}, {Armitage}, {Wood} \& {Lodato}}{{Rice}
  et~al.}{2006}]{Rice2006}
{Rice} W.~K.~M.,  {Armitage} P.~J.,  {Wood} K.,    {Lodato} G.,  2006, \mnras,
  373, 1619

\bibitem[\protect\citeauthoryear{{Ruge}, {Wolf}, {Uribe} \& {Klahr}}{{Ruge}
  et~al.}{2013}]{2013A&A...549A..97R}
{Ruge} J.~P.,  {Wolf} S.,  {Uribe} A.~L.,    {Klahr} H.~H.,  2013, \aap, 549,
  A97

\bibitem[\protect\citeauthoryear{{Ruge}, {Wolf}, {Uribe} \& {Klahr}}{{Ruge}
  et~al.}{2014}]{2014A&A...572L...2R}
{Ruge} J.~P.,  {Wolf} S.,  {Uribe} A.~L.,    {Klahr} H.~H.,  2014, \aap, 572,
  L2

\bibitem[\protect\citeauthoryear{{Sallum}, {Follette}, {Eisner}, {Close},
  {Hinz}, {Kratter}, {Males}, {Skemer}, {Macintosh}, {Tuthill}, {Bailey},
  {Defr{\`e}re}, {Morzinski}, {Rodigas}, {Spalding}, {Vaz} \&
  {Weinberger}}{{Sallum} et~al.}{2015}]{2015Natur.527..342S}
{Sallum} S.,  {Follette} K.~B.,  {Eisner} J.~A.,  {Close} L.~M.,  {Hinz} P.,
  {Kratter} K.,  {Males} J.,  {Skemer} A.,  {Macintosh} B.,  {Tuthill} P.,
  {Bailey} V.,  {Defr{\`e}re} D.,  {Morzinski} K.,  {Rodigas} T.,  {Spalding}
  E.,  {Vaz} A.,    {Weinberger} A.~J.,  2015, \nat, 527, 342

\bibitem[\protect\citeauthoryear{{Shakura} \& {Sunyaev}}{{Shakura} \&
  {Sunyaev}}{1973}]{shakura_1973}
{Shakura} N.~I.,  {Sunyaev} R.~A.,  1973, \aap, 24, 337

\bibitem[\protect\citeauthoryear{{Stone} \& {Norman}}{{Stone} \&
  {Norman}}{1992}]{Zeus}
{Stone} J.~M.,  {Norman} M.~L.,  1992, \apjs, 80, 753

\bibitem[\protect\citeauthoryear{{Takeuchi} \& {Lin}}{{Takeuchi} \&
  {Lin}}{2002}]{TakeuchiLin2002}
{Takeuchi} T.,  {Lin} D.~N.~C.,  2002, \apj, 581, 1344

\bibitem[\protect\citeauthoryear{{Tanaka}, {Takeuchi} \& {Ward}}{{Tanaka}
  et~al.}{2002}]{2002ApJ...565.1257T}
{Tanaka} H.,  {Takeuchi} T.,    {Ward} W.~R.,  2002, \apj, 565, 1257

\bibitem[\protect\citeauthoryear{{Varni{\`e}re}, {Bjorkman}, {Frank},
  {Quillen}, {Carciofi}, {Whitney} \& {Wood}}{{Varni{\`e}re}
  et~al.}{2006}]{varniere20062}
{Varni{\`e}re} P.,  {Bjorkman} J.~E.,  {Frank} A.,  {Quillen} A.~C.,
  {Carciofi} A.~C.,  {Whitney} B.~A.,    {Wood} K.,  2006, \apjl, 637, L125

\bibitem[\protect\citeauthoryear{{Varni{\`e}re}, {Quillen} \&
  {Frank}}{{Varni{\`e}re} et~al.}{2004}]{2004ApJ...612.1152V}
{Varni{\`e}re} P.,  {Quillen} A.~C.,    {Frank} A.,  2004, \apj, 612, 1152

\bibitem[\protect\citeauthoryear{{Wagner}, {Apai}, {Kasper} \&
  {Robberto}}{{Wagner} et~al.}{2015}]{2015ApJ...813L...2W}
{Wagner} K.,  {Apai} D.,  {Kasper} M.,    {Robberto} M.,  2015, \apjl, 813, L2

\bibitem[\protect\citeauthoryear{{Weidenschilling}}{{Weidenschilling}}{1977}]{metersizedproblem}
{Weidenschilling} S.~J.,  1977, \mnras, 180, 57

\bibitem[\protect\citeauthoryear{{Weingartner} \& {Draine}}{{Weingartner} \&
  {Draine}}{2001}]{Weingartner2001}
{Weingartner} J.~C.,  {Draine} B.~T.,  2001, \apj, 548, 296

\bibitem[\protect\citeauthoryear{{Williams} \& {Cieza}}{{Williams} \&
  {Cieza}}{2011}]{WilliamsCieza}
{Williams} J.~P.,  {Cieza} L.~A.,  2011, \araa, 49, 67

\bibitem[\protect\citeauthoryear{{Zhang}, {Bergin}, {Blake}, {Cleeves},
  {Hogerheijde}, {Salinas} \& {Schwarz}}{{Zhang}
  et~al.}{2016}]{2016arXiv160105182Z}
{Zhang} K.,  {Bergin} E.~A.,  {Blake} G.~A.,  {Cleeves} L.~I.,  {Hogerheijde}
  M.,  {Salinas} V.,    {Schwarz} K.~R.,  2016, ArXiv e-prints

\bibitem[\protect\citeauthoryear{{Zhang}, {Blake} \& {Bergin}}{{Zhang}
  et~al.}{2015}]{2015ApJ...806L...7Z}
{Zhang} K.,  {Blake} G.~A.,    {Bergin} E.~A.,  2015, \apjl, 806, L7

\bibitem[\protect\citeauthoryear{{Zhu}, {Nelson}, {Dong}, {Espaillat} \&
  {Hartmann}}{{Zhu} et~al.}{2012}]{Zhu2012}
{Zhu} Z.,  {Nelson} R.~P.,  {Dong} R.,  {Espaillat} C.,    {Hartmann} L.,
  2012, \apj, 755, 6

\bibitem[\protect\citeauthoryear{{Zhu}, {Stone}, {Rafikov} \& {Bai}}{{Zhu}
  et~al.}{2014}]{zhu2014}
{Zhu} Z.,  {Stone} J.~M.,  {Rafikov} R.~R.,    {Bai} X.-n.,  2014, \apj, 785,
  122

\end{thebibliography}
\bibliographystyle{mn2e}

\bsp

\label{lastpage}

\end{document}